\documentclass[lscape,numberedappendix,appendixfloats]{emulateapj}

\usepackage{amsfonts, color, hyperref, longtable}
\usepackage{psfig}
\usepackage{amsmath}
\usepackage{amssymb}
\newcommand {\um}{$\mu$m}

\def\um     {$\mu$m}
\def\ums     {$\mu$m~}
\def\ts     {\thinspace}
\def\kms    {\ifmmode{{\rm \ts km\ts s}^{-1}}\else{\ts km\ts s$^{-1}$}\fi}
\def\msol   {\ifmmode{{\rm M}_{\odot}}\else{M$_{\odot}$}\fi}
\def\lsol   {\ifmmode{{\rm L}_{\odot}}\else{L$_{\odot}$}\fi}
\def\zsol   {\ifmmode{{\rm Z}_{\odot}}\else{Z$_{\odot}$}\fi}

\def\ltsima{$\; \buildrel < \over \sim \;$}
\def\simlt{\lower.5ex\hbox{\ltsima}}
\def\gtsima{$\; \buildrel > \over \sim \;$}
\def\simgt{\lower.5ex\hbox{\gtsima}}

\newcommand{\msun}{{\rm\,M$_\odot$}}
\newcommand{\sfr}{{\rm\,M$_\odot$\,yr$^{-1}$}}

\newcommand{\hst}{\textit{HST}}
\newcommand{\spitzer}{\textit{Spitzer}}
\newcommand{\chandra}{\textit{Chandra}}
\newcommand{\xmm}{\textit{XMM-Newton~}}
\newcommand{\xmms}{\textit{XMM-Newton}}
\newcommand{\ergs}{${\rm erg \ cm^{-2} \ s^{-1}}$ }
\newcommand{\erg}{${\rm erg \ s^{-1}}$ }
\newcommand{\ergsa}{${\rm erg \ cm^{-2} \ s^{-1} \AA^{-1}}$ }
\def\gax{{$\mathrel{\hbox{\rlap{\hbox{\lower4pt\hbox{$\sim$}}}\hbox{$>$}}}$}}
\newcommand{\sersic}{S\'{e}rsic}
\newcommand{\src}{{\rm\,J1302~}}
\newcommand{\srcs}{{\rm\,J1302}}

\def\ltsima{$\; \buildrel < \over \sim \;$}
\def\simlt{\lower.5ex\hbox{\ltsima}}
\def\gtsima{$\; \buildrel > \over \sim \;$}
\def\simgt{\lower.5ex\hbox{\gtsima}}

\shorttitle{J1302}
\shortauthors{Shu et al.}

\begin{document}


\title{Central Engine and Host Galaxy of RXJ 1301.9+2747: A Multi-wavelength view 
of a Low-mass Black Hole Active Galactic Nuclei with Ultrasoft X-ray Emission}
%\title{Multi-wavelengh Study of a Low-mass Black Hole Active Galactic Nuclei with Ultrasoft X-ray Emssion}

\author{X.W. Shu\altaffilmark{1}, 
T. G. Wang\altaffilmark{2}, 
N. Jiang\altaffilmark{2},
J. X. Wang\altaffilmark{2},
L. M. Sun\altaffilmark{2}, 
H. Y. Zhou\altaffilmark{2,3}
}
\altaffiltext{1}{Department of Physics, Anhui Normal University, Wuhu, Anhui, 241000, China; 
xwshu@mail.ahnu.edu.cn}
\altaffiltext{2}{Department of Astronomy, University of Science and Technology of China, Hefei, Anhui 230026, China}
\altaffiltext{3}{Polar Research Institute of China, 451 Jinqiao Road, Shanghai,
200136, China}
%\affil{USTC/AHNU}
%\setcounter{footnote}{15}

%\author{C. D. Biemesderfer\altaffilmark{4,5}}
%\affil{National Optical Astronomy Observatories, Tucson, AZ 85719}
%\email{aastex-help@aas.org}

%\and

%\author{R. J. Hanisch\altaffilmark{5}}
%\affil{Space Telescope Science Institute, Baltimore, MD 21218}


%\altaffiltext{1}{Visiting Astronomer, Cerro Tololo Inter-American Observatory.
%CTIO is operated by AURA, Inc.\ under contract to the National Science
%\altaffiltext{2}{Society of Fellows, Harvard University.}
%\altaffiltext{3}{present address: Center for Astrophysics,
%    60 Garden Street, Cambridge, MA 02138}
%\altaffiltext{4}{Visiting Programmer, Space Telescope Science Institute}
%\altaffiltext{5}{Patron, Alonso's Bar and Grill}


\begin{abstract}
%We present an analysis of multiwavelength observations of the low mass galaxy 
%RXJ 1301.9+2747 in order to study the properties of the active nuclei and the host galaxy. 
%This galaxy extremely soft X-ray emission and 
RXJ 1301.9+2747 is an optically identified very low mass AGN candidate with $M_{\rm BH}\sim1\times10^6M_{\odot}$, 
which shows extremely soft X-ray emission and unusual X-ray variability in the form of short-lived 
flares. 
We present an analysis of multiwavelength observations of RXJ 1301.9+2747 in order to study the 
properties of the active nucleus and its host galaxy. 
The UV-to-X-ray spectrum in the quiescent state can be well and self-consistently described by 
a thermal and a Comptonized emission from accretion disk, with blackbody dominating $\sim$70\% of the X-rays 
in the 0.2-2 keV. 
{The same model can describe the X-ray spectrum in the flare state but the 
Comptonized component becomes dominant ($\sim80$\%). 
The best-fit implies an Eddington ratio of $\sim0.14$ 
and a black hole mass of $1.7-2.8\times10^6$\msun}, in agreement with the estimation 
from the optical data within errors.   
However, the best-fitting model under-predicts the optical flux for the \hst~point source 
by a factor of $\sim$2. 
The excess of nuclear optical emission could be attributed to a nuclear stellar cluster 
which is frequently seen in low mass AGNs. 
The X-ray to optical spectral slope ($\alpha_{\rm ox}$) is lower than in most 
other active galaxies, which may be attributed to intrinsically X-ray 
weakness due to very little hot and optically thin coronal emission. 
%No any weak or hidden broad emission line 
We performed a pilot search for weak or hidden broad emission lines using 
optical spectropolarimetry observations, but no any polarized broad lines 
are detected.  
%RXJ 1301.9+2747 is detected significantly in the radio, but the luminosity is still 
%too weak to be consistent with radio loud, making the origin of the radio emission intriguing 
%if the source is in analog with X-ray binary systems at the high/soft state.  
The host galaxy appears to be a disk galaxy with a boxy pseudobulge or nuclear bar accounting 
for $\sim$15\% of the total starlight, which is consistent with the general characteristics of 
the host of low mass AGNs.   
\end{abstract}

\section{Introduction}
%The origin of 
Supermassive black holes (SMBHs) with masses of $M_{\rm BH}\sim10^6-10^9~M_{\odot}$ 
%via accreting are commonplace in Active galactic nuclei (AGNs), and 
reside in the center of most bulge-dominated galaxies in the local universe. 
The discovery of the correlation between BH masses and the stellar velocity dispersions, 
$\sigma_{\star}$, of their host galaxies \citep[$M_{\rm BH}-\sigma_{\star}$ relation,][] 
{Gebhardt00,Ferrarese00},
strongly suggests coevolution of galaxies and BHs. 
It is however not known when and how this relation is established. 
In current models of galaxy evolution, SMBHs must have formed from much less 
massive ``seed" BHs and grown up by fast accretion and/or merging { \citep[e.g.,][]{white78, volonteri03, 
begelman06}.}
The nature of such seed BHs is thus the major challenge to any cosmological BH 
growth model. 
 
% early in the evolution process.  

%are a ubiquitous component of Active galactic nuclei (AGNs). 
%Though the formation mechanism and growth of "seeds" for these BHs are 
%still poorly understood, the tight correlations of their masses with the properties 
%of host spheroidal component suggest that they play an essential role in the evolution of galaxies. 
%are formed from the collapse of massive stars (e.g., Fryer 2003), 
BHs with masses in the range of $10^4-10^6~M_{\odot}$ (IMBHs hereafter) %presumably hosted by late-type or dwarf galaxies (e.g., van der Marel 2004; Jiang et al. 2011), 
are of particularly important. 
By filling the mass gap 
between supermassive and stellar-mass BHs, 
these objects are potential analogs of the seeds of supermassive BHs 
\citep[e.g.,][]{Volonteri10, greene12}. 
%but also provide an excellent diagnostic tool for probing the BH accretion processes 
%in an underexplored regime of parameter space (Ho et al. 2012; Yuan et al. 2014). 
Recent optical and X-ray observations have yielded a sample of $\sim$200-300 
candidate IMBHs which revealed themselves as active nuclei in small galaxies 
\citep{greene04, greene07b, dong12, kamizasa12, schramm13, reines13, moran14}, 
 %(Greene \& Ho 2004, 2007b; Dong et al. 2012b; Kamizasa et al. 2012; Schramm et al. 2013;  
%Reines et al. 2014; Moran et al. 2014), 
and are beginning to constrain formation and evolution models of seed BHs \citep{greene12}.  
%The two prototypes examples of IMBH AGNs are the late type spiral galaxy NGC 4395 
%(Filippenko \& Ho 2003) and the dwarf elliptical galaxy POX 52 (Barth et al. 2004).
In fact, deviations in the $M_{\rm BH}-\sigma_{\star}$ relation have been observed 
in a population of IMBHs \citep[e.g., see review by][]{kormendy13}. 
%that is not yet fully grown (e.g., Xiao et al. 2011). 
The host galaxies of IMBHs appear very different from their supermassive counterparts.
Detailed bulge-disk-bar decompositions of IMBH AGNs  
%of a large sample of low-mass BH AGNs find that the 
have shown that majority of galaxies have disks and are likely to contain pseudobulges, with very few of 
them living in classical bulges \citep{greene08, jiang11a}
%(Greene et al. 2008; Jiang et al. 2011a), 
suggesting 
that bulge may not be a necessary condition for BH growth and 
a secular evolution of the low mass BHs is favored. 

Presumably, IMBHs likely have not had the opportunity to become full-grown. 
The accretion process and the disk corona geometry around the central IMBH could be
markedly different from that around SMBHs in luminous AGNs 
{\citep[e.g.,][]{godet12, miniutti13, jin16}}.
Therefore, AGNs with IMBHs provide an excellent diagnostic tool for probing the
BH accretion processes in a {poorly explored regime of parameter space 
\citep[e.g.,][]{dewangan08, miniutti09, yuan10, dong12, yuan14, pan15, ludlam15, ho16, plotkin16}}. 
%They also provide the missing link between the stellar mass and SMBHs and enable us 
They are potential candidates enabling us 
to test if the accretion physics is the same at all scales from 
the stellar mass to supermassive BHs {\citep[][]{mchardy06, gultekin14, zhou15}}. 
%However, \chandra~and \xmm observations of IMBH AGNs discovered with SDSS have shown 
%that the X-ray spectral properties are consistent with 
%that for luminous AGNs, implying no obvious dependence on BH masses  
%(Greene \& Ho 2007a; Desroches et al. 2009; Miniutti et al. 2009). 
%Gultekin et al. (2014) find that low mass BHs do belong
%the "fundamental plane of BH accretion", a correlation between the continuum
%X-ray and radio emission and mass of accreting supermassive and 
%stellar-mass BHs (e.g., Gallo et al. 2012).

% In fact, very little is known about the accretion process onto IMBHs in AGNs.  
%Therefore, it is crucial 
%to study accretion onto the least massive central BHs in galaxies. 
%To date a number of 

%Summary of 
%There is a gap of 5 orders of magnitude in BH mass between 
%stellar-mass BHs with masses of $\sim$10$M_{\odot}$ (formed from 
%the death of massive stars) and Supermassive BHs. 
%The recent discoveries of previously unknown galactic BHs are beginning to constrain formation and evolution models of such seed BHs. NGC 4395, 

The broadband spectral energy distribution (SED) of AGNs are powerful diagnostics on the 
detailed BH accretion processes as well as the interplay with the host galaxies 
\citep[e.g.,][]{elvis94, ho99, vasudevan09, jin12a}.
%(e.g., Elvis et al. 1994; Ho 1999; Vasudevan et al. 2009; Jin et al. 2012). 
Although it is not straightforward to measure the SEDs for low mass and hence low luminosity 
AGNs because of the strong contamination from the hosts, 
there are increasing observational evidence suggesting the unusual multiband properties for these systems. 
%Low mass AGNs with high accretion rates have been found very radio-quiet on average 
%(Greene et al. 2006). % and maybe intrinsically X-ray weak (Dong et al. 2012; Yuan et al. 2014). 
\chandra~and \xmm observations of IMBH AGNs discovered with SDSS have shown 
%that the sources are relatively X-ray bright, though  
that while the soft X-ray spectral properties are consistent with that for AGNs with 
more massive BHs, 
{ the X-ray-to-optical spectral slopes ($\alpha_{\rm ox}$) tend 
to be steeper (on average) than expected for their UV luminosities,}
%on average the X-ray-to-optical spectral slope ($\alpha_{\rm ox}$) tends to lie below the low-luminosity extrapolation of the well-known $\alpha_{\rm ox}-
%L_{2500\AA}$ correlation, }
%the optical-to-X-ray SEDs 
%the X-ray-to-optical spectral slope is systematically steeper, 
suggesting that at least some of them may be intrinsically X-ray weak 
\citep{greene07a, desroches09, miniutti09, dong12}.
%(Greene \& Ho 2007a; Desroches et al. 2009; Miniutti et al. 2009; Dong et al. 2012). 
Low mass AGNs with high accretion rates ($L_{\rm bol}/L_{\rm Edd}\simgt0.1$) have also been found very radio-quiet on average \citep{greene06}. 
%Gultekin et al. (2014) have shown that low mass BHs do belong 
%the ``fundamental plane of BH accretion", a correlation between the continuum 
%X-ray and radio emission and mass of accreting supermassive and stellar-mass BHs (Falcke et al. 2004; 
%Gallo et al. 2012). 
%On the other hand, the X-ray emission of relatively low Eddington ratio sources 
%might be intrinsically weak,  
%By extending the study a new sample of relatively low Eddington ratio sources, 
%Yuan et al. (2014) 
\citet{yuan14} recently extended the study to relatively low Eddington ratio sources 
($L_{\rm bol}/L_{\rm Edd}\sim10^{-2}$, {Plotkin et al. 2016}), and found that they tend to be intrinsically 
X-ray weak, very similar to that of the prototype IMBH AGN NGC 4395. 
On the other hand, \citet[][]{gultekin14} have shown that low mass BHs do belong         
the ``fundamental plane of BH accretion", a correlation between the 
X-ray, radio continuum emission and mass of accreting BHs 
{\citep{merloni03, falcke04, gallo12}}.

More recently, \citet{terashima12} reported an extreme case of IMBH AGN, 
%An extreme case of such object is found recently, 
2XMM J123103.2+110648 (J1231+1106), 
with a central BH mass as small as $10^5M_{\odot}$ (Ho et al. 2012), 
whose X-ray spectrum completely lacks emission at energies
$\simgt2$keV, and can be described entirely by a soft thermal component of 
temperature $kT\sim0.12$ keV.  
Such an extreme soft excess is unprecedented among AGNs, even at the very low mass end 
\citep[e.g.,][]{dewangan08, desroches09, miniutti09, ai11}. 
%(e.g., Dewangan et al. 2008; Desroches et al. 2009; Miniutti et al. 2009; Ai et al. 2011). 
A similarly extremely soft X-ray spectrum was also found in GSN 069 ($M_{\rm BH}\sim1.2\times10^6M_{\odot}$) 
by \citet{miniutti13}.
% for GSN 069,  
%a source without detection of broad $H\alpha$ or $H\beta$ lines as in J1231+1106. 
%Miniutti et al. attribute 
%a source showing many similarities as J1231+1106.;
In either case, the $pure$ thermal X-ray spectrum appears to be a close analog to the typical high/soft state in BH X-ray binaries (BHBs), 
which would suggest a new, and possibly accretion disk dominated, AGN spectral state.
The discovery of a $\sim$3.8 hr periodicity from J1231+1106 adds 
further confidence of the similarity of this ultrasoft AGN to the BHBs \citep{lin13}.
%However, while the UV-to-X-ray spectrum of GSN 069 is consistent with a pure accretion 
%disk model, the spectrum of J1231+1106 can be described with either pure thermal disk 
%emission or optically thick low-temperature Comptonization.
%If it is thermal, the blackbody temperature  
%is higher than that predicted by the standard accretion disk model. 
\citet{lin13} claimed that the ultrasoft X-ray spectra in the two objects 
could be associated with tidal disruption events, as both clearly show long-term luminosity evolution, 
making the nature of their ultrasoft X-ray emission intriguing. 
{RX J1301.9+2746 (hereafter J1302) is another ultra-soft and highly variable 
AGN (Sun, Shu \& Wang 2013, hereafter S13).
Our detailed analysis of the optical spectrum taken from the SDSS
revealed that the galaxy hosts a Seyfert-like nucleus at $z=0.0237$. 
Using the width of the [OIII]$\lambda$5007 line as a proxy for the stellar velocity dispersion of the host galaxy,
we obtained a BH mass of $M_{\rm BH}\sim8\times10^5M_{\odot}$ 
with an intrinsic scatter of 0.5 dex, placing
J1302 in the regime of IMBHs.}
The X-ray properties of this source not only show many similarities to J1231+1106 and GSN069 
(lack hard X-rays and no detectable broad H$\alpha$ and H$\beta$ lines), 
but also exceptional in at least two respects: 
1) {Its X-ray light curve shows clearly two distinct states: 
a long quiescent state and a short flare (or eruptive) state},
which differs in count rates by a factor of 5--7; 
%2) {The X-ray spectrum of J1302 is extreme steep ($\Gamma>7$ in the quiescent state) 
%which is unprecedented among AGNs;} 
2) Significant detection in the radio (1.4 GHz) with VLA ($\sim$7$\sigma$). 
%Given the apparent radio quiescence of low mass AGNs found by \citet{greene06},
%J1302 is somewhat unusual in its radio emission. 
{Because of these unique properties, J1302 remains a valuable target for further study. }
Here we present an analysis of multi-wavelength observations of J1302, including X-ray from 
\chandra/\xmms, optical from HST, ultraviolet (UV) from \xmms/OM and GALEX, near-to-mid infrared from 2MASS, 
WISE and \spitzer, and radio from the Very Large Array (VLA), 
in order to investigate the nuclear SED of J1302 over a wide range in frequency as well as 
the host galaxy properties. In Section 2, we present the observations and data analysis. 
The results are discussed and concluded in Section 3 \& 4, respectively. 
%The galaxy J1302, also known as KUG 1259+280, is a post-starburst galaxy in the Coma cluster. 
% appears to be in a small group of four galaxies within an arcminute. 
%From ground-based imaging, J1302 was classified as an S0 galaxy. 
%However, the high resolution {\it Hubble Space Telescope} observations have found this 
%galaxy to be an edge-on disk galaxy with a very bright unresolved nucleus (Caldwell, Rose \& Dendy 1999). 
%Given the radio quiescence of nearly all low-mass AGNs studied in Greene \& Ho (2006), 

%In this paper, 
%In their deep VLA 1.4 GHz survey of the Coma cluster, Miller et al. 2009 have found the 
%J1302 was associated with a radio source
%Surprisingly, J1302 was detected at 1.4 GHz 

%of POX 52, including the first targeted HST and X-ray observations of this unusual object. HST images are used to examine the host galaxy morphology; 
%HST nuclear SED.
 
\section{Observations and data}

\subsection{Spectropolarimetric Observation}

{Spectropolarimetric observation was made with the CCD
Spectropolarimeter (SPOL; Schmidt et al. 1992) at the 6.5 m Multiple Mirror Telescope (MMT) 
on 2016 April 3. 
We used a low-resolution grating (6001 mm$^{-1}$) providing spectral coverage of 4100--8200\AA. 
An entrance slit of $\sim$1.\arcsec1 width centered on the nucleus was used to match the seeing. 
%The total exposure time was 40 minutes 
The resulting spectral resolution is $\sim$15\AA~FWHM, 
corresponding to FWHM$\sim700$ km/s around the $H\alpha$. 
The total exposure time was 40 minutes. 
%The observation used 
A $\lambda$/4 rotatable achromatic retardation plate was used to provide 
linear polarization measurements. 
The data reductions, including bias subtraction, flat-field correction, 
and cosmic-ray removal, were accomplished with standard procedures 
using the IRAF script provided by the instrument's PI Dr. Paul Smith
\footnote{http://james.as.arizona.edu/~psmith/SPOL/spolred.html}.  
%The data reduction was done using an IRAF script. (此处用note，http://james.as.arizona.edu/~psmith/SPOL/spolred.html，we thank Paul Smith for sharing us the script）
Stokes $Q$ and $U$ parameters were measured individually 
and then combined to obtain the degree of polarization.
}

\subsection{Analysis of the HST Images}
J1302, as a member of Coma cluster, was observed by \hst/WFPC2 in F814W 
(roughly the $I$ band of the Johnson system) and F450W ($B$) filter 
in a project studying current starburst and post-starburst
early-type galaxies in nearby clusters of galaxies on 1997 July 12 
{\citep[][]{caldwell99}.
Some results from \hst~observations have been presented by \citet[][]{caldwell99}, 
which reveal J1302 to be an edge-on disk galaxy with a very bright nucleus. 
Here we examine the \hst~data in detail in the context of the broadband SED analysis on 
the nucleus.}
For ease of rejecting cosmic rays, the total exposure time
(1200s and 800s for $B$ and $I$ bands respectively) were divided into two 
equal exposures in the observations of each band. 
These images are then combined using {\tt astrodrizzle} to 
remove cosmic-ray hits and to correct for possible geometric distortion.
J1302 was located at the center of PC1 chip and 
thus the final pixel scale is 0.\arcsec0455.

Then we try to perform a two-dimensional (2D) decomposition of J1302 using
GALFIT \citep{peng02, peng10}. The AGN is represented by
a point source modeled with the TinyTim PSF, and the host galaxy is modeled by
a bulge as \sersic\ $r^{1/n}$ function or a disk as exponential (Exp)
function (equivalent to $n = 1$), or a combination of them.
The sky background have been subtracted during {\tt astrodrizzle} 
combination, also confirmed by our residual sky estimation. 
Other objects, like background galaxies or foreground stars are masked out. 
{The single \sersic\ fitting scheme gives an unacceptable large residuals in the image,  
while \sersic\ + Exp fitting scheme yields a much better result. 
For example, in our fittings to the \hst~I-band image, the single \sersic\ model yields fit statistics 
$\chi^2/dof=221594/361180$, while it is significantly improved using the \sersic\ + Exp model 
with $\Delta\chi^2=14804$ for four extra parameters.}
During the B-band fitting, we have fixed the \sersic\ index to be the value 
given by $I$ band because a totally free fitting yields an unreasonable high index.
%The fitting results are summarized in Figure~1 and Table~1.

\begin{figure}
\centering
{
\includegraphics[width=0.8\textwidth]{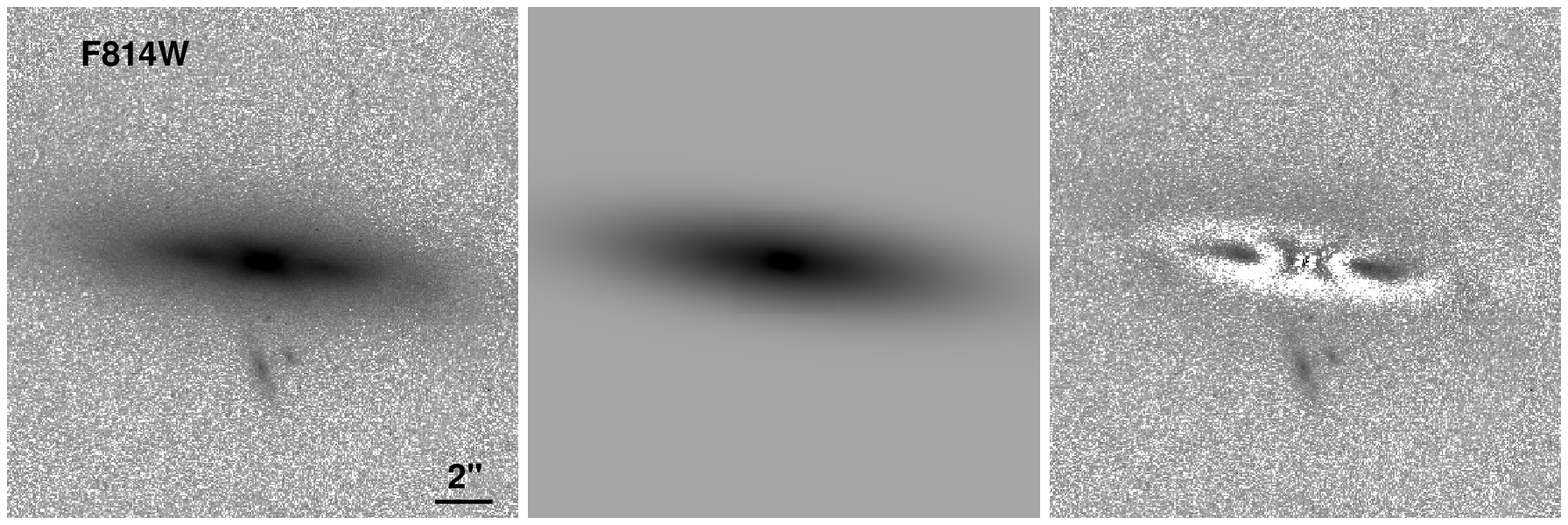}%
\hfil
\includegraphics[width=0.8\textwidth]{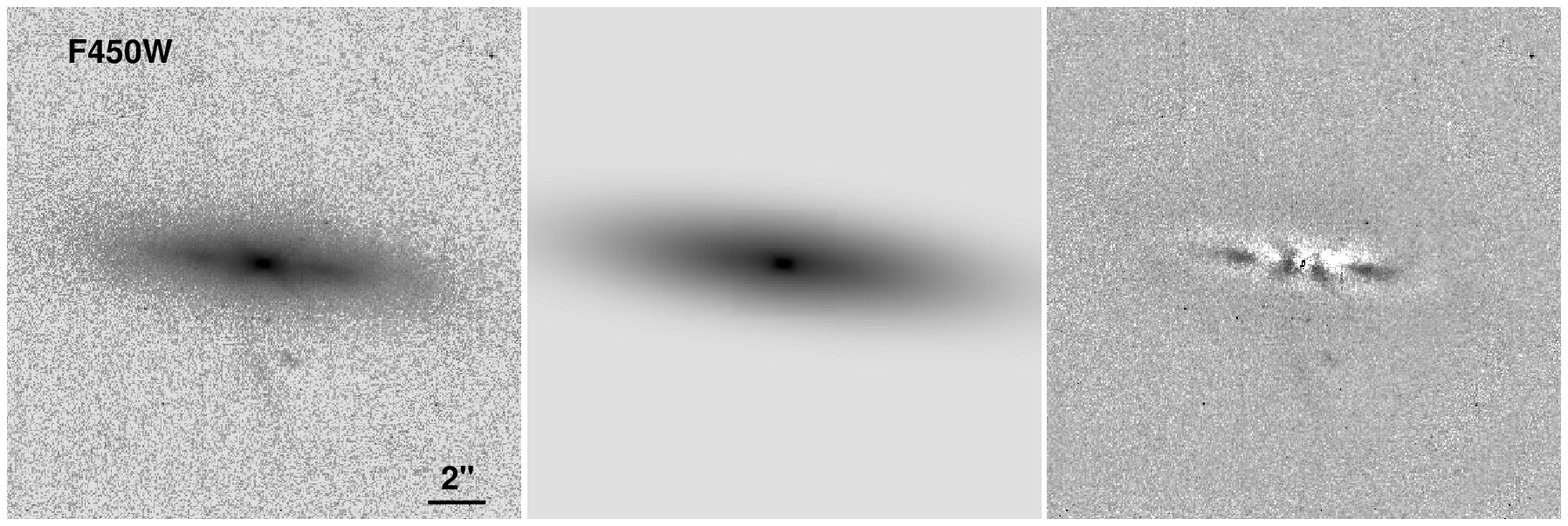}%
\hfil
\includegraphics[width=0.5\textwidth]{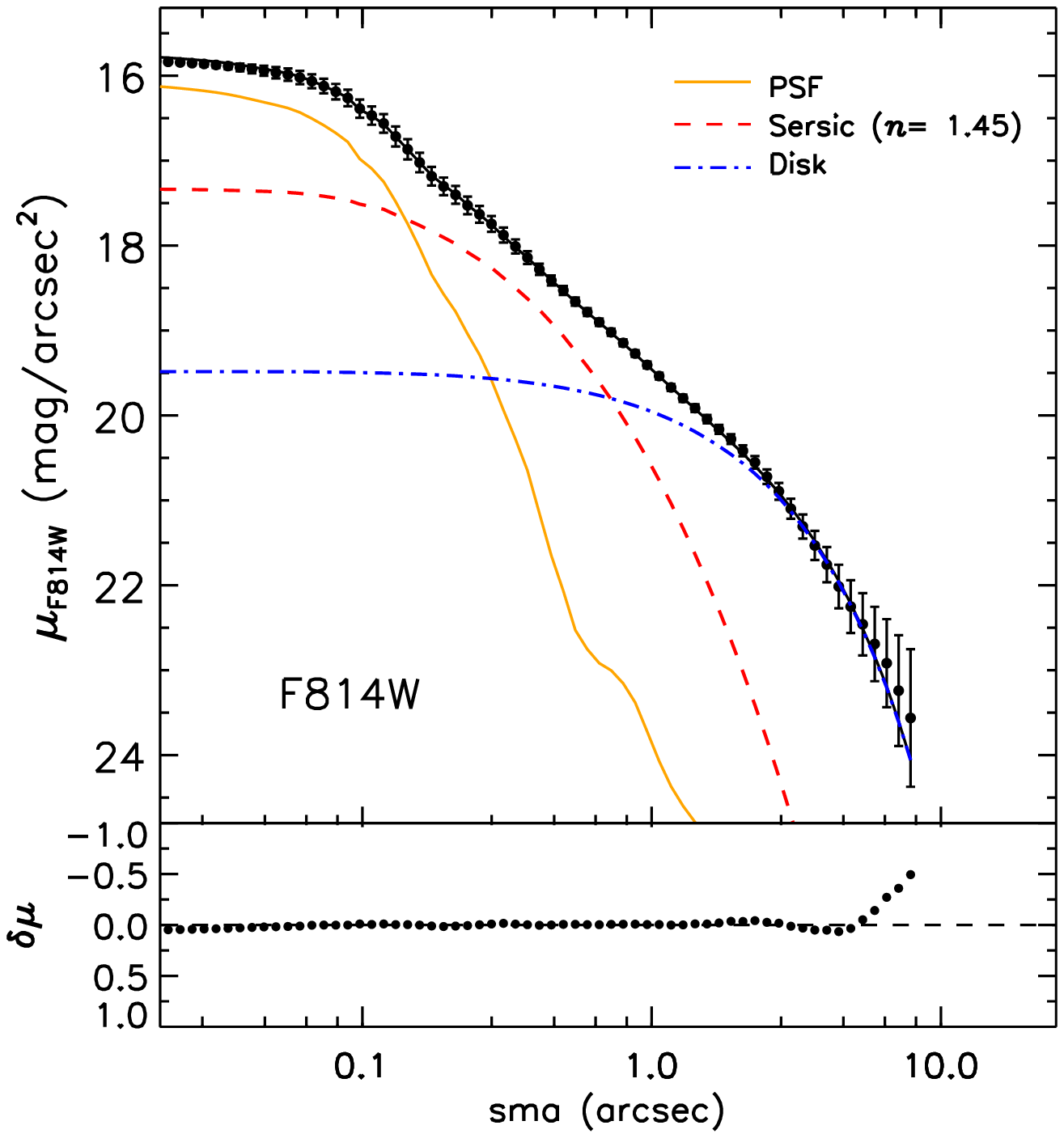}%
\includegraphics[width=0.5\textwidth]{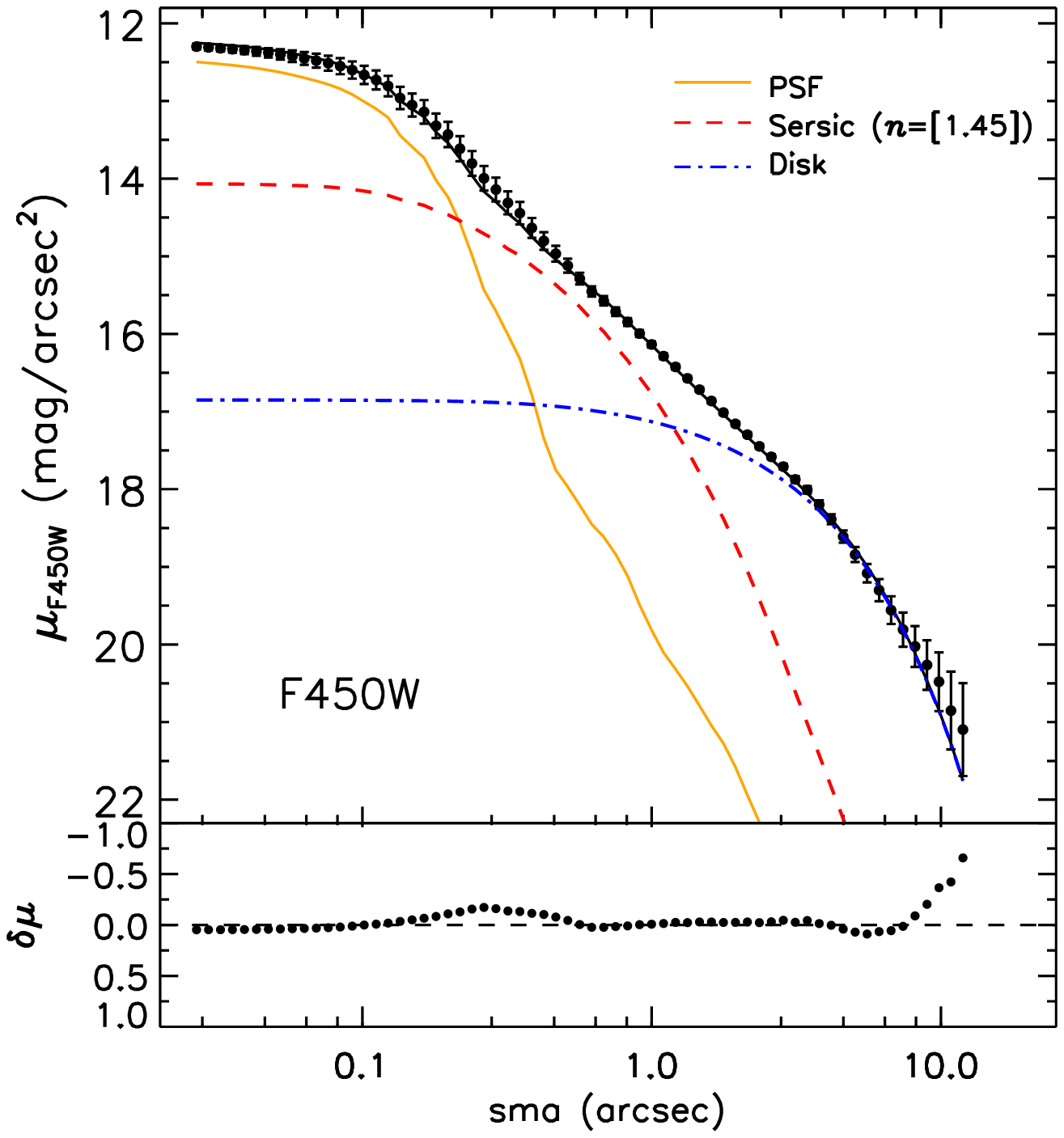}%
%\hfil
}
\caption{
Galfit decomposition of the \hst/WFPC2 F814W and F450W images of J1302.
Top two panels: from left to right shows original image, 
GALFIT model (PSF + \sersic\ + Exp) and the residual image, respectively.
Bottom two panels: one-dimensional representation of the GALFIT fitting,
PSF for the nucleus (orange line), 
$n=1.45$ \sersic\ function for the bulge (red dashed line), and
an exponential function for the disk (blue dot-dashed line).  The sum of
the three components is shown as the black solid line. The observed data are
plotted as black symbols with $\pm 1 \sigma$ error bars.  
}\label{fig:hst_image}
\end{figure}

{ The top panels of Figure~\ref{fig:hst_image} show the F814W and F450W\hst~image (left panel) 
and the GALFIT model (middle panel). {The best-fit yields 
$m_{\rm AGN}=18.1$ ($f_{\lambda}=6.4\times10^{-17}$ \ergs\AA), $m_{\rm bulge}=16.4$ 
($f_{\lambda}=3.16\times10^{-16}$ \ergs\AA) with an effective radius of 0.\arcsec41, 
and $m_{\rm AGN}=19.0$ ($f_{\lambda}=1.37\times10^{-16}$ \ergs\AA), $m_{\rm bulge}=17.6$ 
($f_{\lambda}=5.29\times10^{-16}$ \ergs\AA) }with an effective radius of 0.\arcsec37, 
for the F814W and F450W, respectively. The residuals after
subtracting the model from the original image are shown in the
right panel. The weak residuals seen in the central region ($\sim0.15-0.3$ arcsec), 
especially in the F450W \hst~image,  
could indicate small-scale structure in the host galaxy (Section 3.3).   
The bottom panels of Figure~\ref{fig:hst_image} show a radial profile plot of the 
best fit from GALFIT consisting of a central PSF (orange), a \sersic~ of $n=1.45$ 
(red dashed), and an outer disk (blue dot-dashed). 
{The residuals at large radii ($r>5$ arcsec) are likely due to more extended disk emission 
which deviates from the adopted exponential profile ($n=1$). 
Allowing the \sersic~index of the disk component to vary yields very similar results, 
indicating that the current \hst~observations are not sensitive to detect more extended, 
low surface brightness emission from the galaxy. 
In fact, previous studies have shown that the disk component is always fitted by an exponential profile,
as in nearby inactive galaxies and galaxies with IMBHs, although disk profiles do vary at large radii 
\citep[e.g.,][]{jiang11b}}. 
Parameters of the F814W and F435W fits can be seen in Table 1, which 
are consistent with each other except the magnitudes.   
{Additional checks on the photometry and potential systematic errors are presented in the Appendix A. 
}
%The strong residuals in the center of the image are
}

%The \sersic\ component, identified as pseudobulge usually, is likely
%an edge-on peanut-shaped bar, contributing $\sim15\%$ of the total flux of
%the host galaxy.
%t's noteworthy that there is a significant vertical "X"-shaped structure 
%in the residual image, which is unrare for Milky Way mass galaxies 
%in the local Universe (Laurikainen et al. 2014) and also confirmed 
%by N-body simulation as a natural evolution result 
%from a pure disk galaxy (Li \& Shen 2012). 

\subsection{X-ray Observations}

J1302 was observed by \xmm EPIC cameras on December 2000 with a total exposure time of 29 ks, 
and \chandra~on June 2009 for about 5 ks.
%The detailed descriptions on the data process and 
The data were processed following the standard criteria, which is detailed
in S13.
{ 
%We found unusual giant flares in both \xmm and \chandra light curves, accompanied by
%spectra hardening during the flare state. 
We found in both \xmm and \chandra~observations that the source 
displays ultrasoft X-ray emission and unusual giant flares in the light curves 
for a duration of $\sim2$ ks. 
Though only one and a possible decline of another flare are recorded in the 
\xmm observation, a very similar flare seen in the \chandra~data is suggesting 
that the flare itself appears repetitive and occurs very frequently in
the object.  
%clearly displays distinct states in the X-ray band: a long quiescent state and a short flare (or eruptive) 
%state which differs in count rates by a factor of 5--7.
In this paper, we will present a more physical description of the X-ray spectrum of J1302,
by jointly fitting the X-ray and UV data (see Section 3.1). 
We will use principally the \xmm PN data, which have much higher sensitivity. 
As the source shows peculiar temporal and spectral behaviors, we attempted to quantify 
the spectral variability during flares by dividing the data into high and low flux
intervals, using count rate thresholds of 0.35 counts s$^{-1}$ for the PN data. 
We classify the data above the count rate thresholds as belonging to
the flare state, and those which fall below, to the quiescent state. 
%We will present a more physical description of the X-ray spectrum of J1302,
%by jointly fitting the X-ray and UV data (see Section 3.1).
}

\subsection{Ultraviolet Observations}
\subsubsection{\xmm Optical Monitor}

The Optical Monitor (OM) data from \xmm for the J1302 were taken simultaneously with the X-ray 
observations, using the UVW2 and UVW1 filter, which is centered at 2120\AA~ and 
2910\AA, respectively. The FWHM of the PSF is $\sim$2\arcsec, or 2.1
pixels.  
%The FWHM of the OM PSF is $\sim1.8$\arcsec.  
However, only a few OM images {(four)} were taken and hence we are not able to construct the 
UV lightcurve to examine whether the UV flux is variable, i.e., significant 
flux enhancements as seen in the X-ray. 
Top panels of Figure~\ref{fig:uv_image} show the \xmms/OM UVW2 (left)
and UVW1 images (right).  
Inspection of the images shows that the source is extended, in
particular in the UVW1 image. 
In order to determine the AGN and host galaxy contribution quantitatively, 
we first tried to perform 2D decomposition using GALFIT for the UVW1 filter, 
which is close to the \hst~F450W band. 
However, as the OM UV spatial resolution is worse than the HST, we
choose a simple model consisting of an AGN point source and a single
\sersic~component. 
The \sersic~index of the host galaxy component is slightly larger than in
the \hst~filters with $n=2.06$. 
The best-fit effective radius of the host galaxy, as described by the
\sersic~component, is also larger than
in the F450W filter (3.\arcsec5 compared to 0.\arcsec37), 
and the PSF component has a relatively low luminosity. 
Note that excluding the PSF component results in acceptably residuals, but 
the \sersic~index is a factor of two higher. 
This is possibly because the single \sersic~ component tend to trace the
concentrated light in the PSF-dominated region which is supported by
the fitted smaller effective radius of 2.\arcsec75. 

{Details on the tests on the AGN and host galaxy decomposition with simulations are
presented in Appendix. 
%The resulting UV fluxes for point source (AGN) and for galaxy are presented in Table 2.  
The flux density from the best-fit (PSF+\sersic) model is
$f_\lambda(2120\AA)=1.45\times10^{-17}$\ergsa for the AGN and
$f_\lambda(2120\AA)=6.45\times10^{-16}$\ergs$\AA^{-1}$ for the galaxy,
respectively.  
Note that given the poor resolution, the PSF component likely consists of 
some host galaxy light which is difficult to be disentangled. 
%Therefore the PSF magnitude may describe an upper limit to the magnitude 
%of the AGN. 
We also performed the AGN and host galaxy modeling in the UVW2, but
with parameters for the \sersic~component held fixed except its
magnitude. 
The results for the UVW2 fits can be seen in Table A1 in the Appendix. }
%images are shown in ???
%We used the {\it omichain} routin to measure the source and construct a UV light curve, shown in
%Figure XXX. 
%The measure flux was corrected for Galactic extinction by using the maps of Schlegel et al. (1998) 
%and the reddening curve of ???
%The flux density is $f_\nu(2310\AA)=??$ and $f_\nu(2310\AA)=??$. 
%The light curve shows that the source is not highly variable in the UV. 
%The observed variability is therefore generally consistent with the expected level of variations 
%from photon- counting statistics for a constant source. 

\subsubsection{GALEX} 
\begin{figure}
\centering{
\includegraphics[scale=0.6, angle=0]{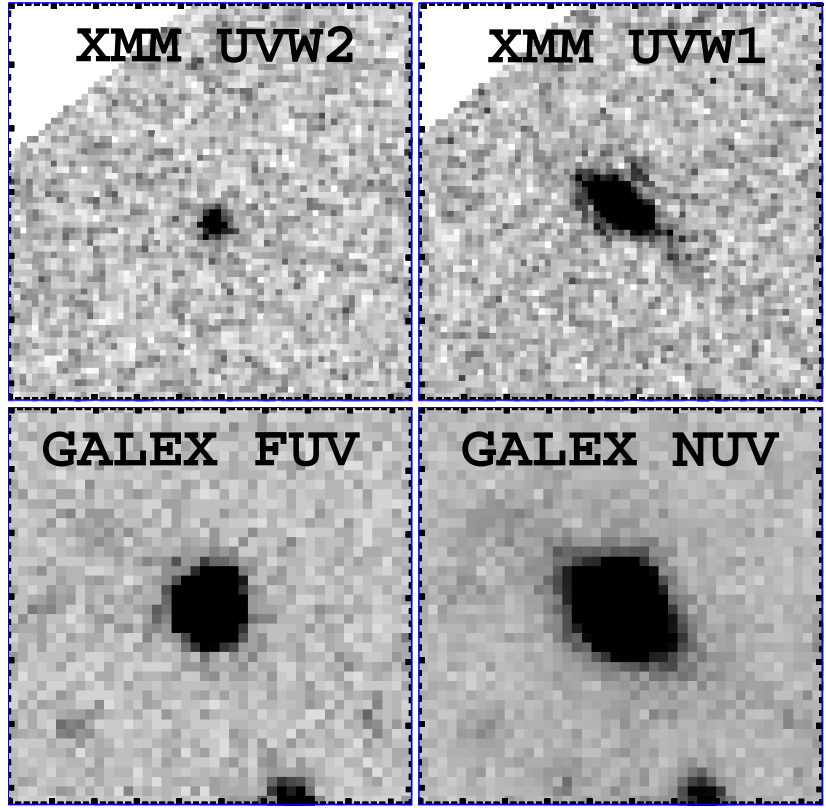}
\caption{
UV images of J1302. The image in the upper row takes from \xmm/OM 
at wavelength $\lambda_{\rm eff}=2120$\AA (UVW2) and $\lambda_{\rm eff}=2910$\AA (UVW1), 
while lower row shows, from left to right, the image in GALEX FUV (1539\AA) and NUV 
(2316\AA), respectively. 
%It is clear that the galaxy in the shorter UV wavelength is dominated by a point source 
%emission, possibly associated with an AGN.  
Each image has a size of $1\times1$ arcmin$^2$.  
%The vertical dotted lines correspond to the FWHM of H$\beta$ line (Wang et al. 2007).}
}\label{fig:uv_image}}
\end{figure}

{Beside the \xmm OM data, J1302 was also imaged and detected in the near-UV (NUV; $\lambda_c=$2316\AA) and 
far-UV (FUV; $\lambda_c=$1539\AA) by GALEX All-sky Imaging Survey on April 2009, for a total exposure time of  
$\sim$26 ks and $\sim$18 ks, respectively.  
The FWHM of the PSF for the NUV filter is 4.\arcsec0 ($\sim$2-3 pixels) 
and for the FUV filter is 5.\arcsec6~($\sim$3-4 pixels). 
%J1302 was also observed during ?? on, for a tal of ??
%Galactic extinction was calculated using the same method as for the OM, and was 
%then used to correct the fluxes taken from the AIS catalog.
% resulting in an NUV flux of $f_\nu(2316\AA)=3.78\times10^{-17}$\ergsa and an FUV flux of $f_\nu(1539\AA)=2.57\times10^{-16}$\ergsa. 
%The NUV band of GALEX surrounds the UVM2 filter on the OM, giving us the opportunity to check for long-term variability in the UV. The flux densities calculated from both instruments are consistent with 
%each other within $\sim$3.3, meaning that there is no significant variability in UV between the two 
%separate observations. 
%Figure~\ref{fig:uv_image} shows the \xmm OM images of  NUV and FUV imagesone 
%taken with the OM on XMM-Newton.
The lower panels of Figure~\ref{fig:uv_image} show 
the GALEX NUV and FUV images, 
{which have not been analyzed before}. 
%Note that the PSF FWHM of \xmm OM filter is $\sim1.8$\arcsec, which is a factor of 
%$\sim$2-3 better than the GALEX (4.0\arcsec~for NUV filter and 5.6\arcsec~for the FUV filter, respectively). 
%It can be seen from the figure that the central AGN appears dominating the emission in the UV, 
%especially in the shorter wavelengths (\xmms/OM UVM2 and GALEX FUV) where the emission is 
% point-like.
%In order to determine the UV host galaxy contribution quantitatively, 
As in the \xmms/OM, the source is extended in the both images.
We tried to perform similar 2D decomposition of the AGN and galaxy emission 
using GALFIT. 
Our GALFIT results show that the PSF component has a flux of $f_\lambda(2316\AA)=3.78\times10^{-17}$\ergsa 
and $f_\lambda(1539\AA)=1.06\times10^{-16}$\ergsa, and the \sersic~ has a flux of 
$f_\lambda(2316\AA)=6.39\times10^{-16}$\ergsa  
and $f_\lambda(1539\AA)=3.58\times10^{-16}$\ergsa, respectively.
%The resulting UV fluxes for point source (AGN) and for galaxy are presented in Table 2.  
Since the NUV band of GALEX surrounds the OM UVW2 filter, the flux gives us the opportunity to 
check for long-term variability in the UV. We find that the flux density derived from {\it XMM-Newton}/OM UVW2 is a factor of $\sim$9 higher than that from GALEX NUV for the point source,
meaning that there is a variability for the \src nucleus in the UV. 
However, it should be noted that the GALEX PSF component may consist of
 considerable host galaxy light due to its poorer resolution ($\sim$5\arcsec, or 2.4 kpc), and  
vice versa. 
In this case, the GALFIT decomposition is very uncertain and one has to treat 
the results with caution. 
% are consistent with each other within $\sim$3.3, meaning that there is no significant variability in UV between the two separate observations.  
%We found that even for the highest estimation, the host galaxy UV flux accounting for ?? 
%of the total UV light, confirming that the source is very AGN dominated in the UV.  
%?? of the total,
}

\subsection{Radio observations}
\begin{figure}
\centering{
\includegraphics[scale=0.7, angle=0]{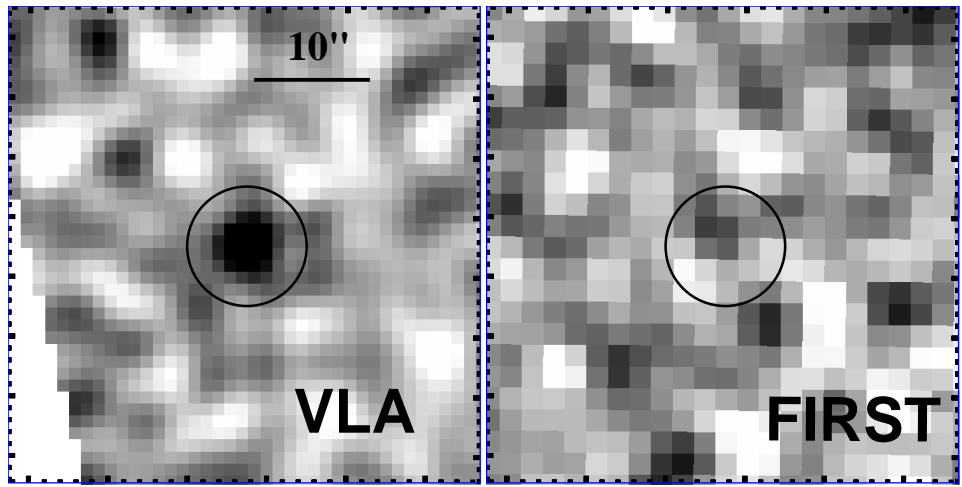}
\caption{
Radio maps from the VLA Coma cluster survey (left, Miller et al. 2009) and FIRST.
The $5\sigma$ detection limit ($<950\mu$Jy, including clean bias) of J1302 from the FIRST survey
is consistent with flux $S_{\rm 1.4 GHz}=779\pm106\mu$Jy measured from the VLA Coma survey.
No apparent evidence of radio variability can be seen from the two existing VLA observations.
The circle has size of 5\arcsec radius.
%The vertical dotted lines correspond to the FWHM of H$\beta$ line (Wang et al. 2007).}
}\label{fig:radio}}
\end{figure}

As part of deep VLA 1.4 GHz imaging of Coma cluster, 
J1302 was observed with VLA on 2006 June. 
The VLA observations were performed in its B configuration, reaching 22 $\mu$Jy 
in the deepest part, or $L_{1.4GHz}\sim1.3\times10^{20}$ W Hz$^{-1}$ for galaxies at the distance of Coma \citep{miller09}. 
%Because of the varying sensitivity across the full mosaic image, 
%source detections were performed in the signal-to-noise ratio (S/N) maps 
%via the task ``Search and Destroy", 
%which directed to identify sources with peaks greater than 4.5$\sigma$ and fit them with Gaussians.
As shown in Figure~\ref{fig:radio}, J1302 was near the edge of the mosaic ($\sim$20\arcsec~from the edge, 
%where the local rms measurement was based on only few resolution elements),
 where the local noise level is high), 
and detected at a level of $\sim7\sigma$. 
%J1302 was not detected in the VLA FIRST survey at 20cm, with a $5\sigma$ limiting flux of
%950 $\mu$Jy/beam including CLEAN bias (Fig3. right).
%Surprisingly, the source was detected at $\sim7\sigma$ in the deeper VLA Coma cluster
%survey at the same frequency (Fig3, left; Miller et al. 2009). 
The peak and integrated 1.4 GHz
flux density is 779$\pm$106 and 866$\pm$197 $\mu$Jy 
{\citep[][]{miller09}, 
corresponding to radio luminosity at 1.4 GHz of $\nu L_{\rm \nu}\sim1.6\times10^{37}$ erg/s. 
}
With the same configuration, J1302 was not detected by Faint Images of the Radio Sky 
at Twenty cm (FIRST) using VLA, with a $5\sigma$ limiting flux of
950 $\mu$Jy/beam (Figure~\ref{fig:radio}, right).
This upper limit is consistent with the above reported flux by \citet{miller09}, 
thus it is not apparent whether
the source is variable or not.
% both are consistent with the FIRST detection limit.
%However, the source was within $\sim$20" of the edge of the VLA Coma survey mosaic, where the
%local rms measurement was based on few resolution elements, leading to the marginal detection.
%It is not apparent from the comparison of two existing VLA observations whether
%the source is variable or not.

With a FWHM of $\sim$5\arcsec, VLA observations cannot resolve the radio emission,
preventing further studies on the origin of it in J1302. 
However, as we discussed below, the radio emission from J1302 is likely associated with 
an AGN, and couldn't be produced by a single BHB, as the highest previous known
is around {$\nu P_{\nu}\sim10^{33}$erg/s \citep[][]{corbel13}}, fare less than what is observed. A young radio
supernova remnant can also be ruled out {since the VLA flux density at 1.4 GHz 
has no evidence of being declined in more than a decade (from 1995 to 2006) 
which would be otherwise detected in the FIRST survey}. The origin of the radio emission from
circumnuclear star formation is also impossible. 
J1302 was detected in 24$\mu$m with a flux of 1.01 mJy (see Section 2.5).
Based on the local galaxy templates \citep{ce01}, the predicted FIR luminosity is $\sim10^9L_{\odot}$.
In combination with the radio power, the FIR/radio ratio is estimated to be $\sim1.6$,
much less than the typical value of 2.6 for star-forming galaxies \citep{ivison10},
suggesting a significant radio excess from AGN.

%Adopting the customary radio-loudness parameter definition, R, as the ratio of the flux densities
%between 6 cm and optical 4400 \AA, 
{We obtained a radio-loudness parameter, $R\sim6$, which is defined as 
the ratio of the flux densities between 6 cm and optical 4400 \AA. 
The latter is the flux for the nuclear point sources as derived from the 
GALFIT-decomposition of \hst~B-band image, 
while the radio flux at 6 cm is estimated from the observed 1.4 GHz flux by assuming a radio 
spectral index $\alpha=0.8$ ($f_{\nu}\sim\nu^{-\alpha}$). 
% parameter for J1302 of $\sim$6  (assuming
%a radio spectral index -0.8).  a radio-loud object , suggesting it is formally radio quiet (Kellermann et al. 1989).
Thus J1302 is formally radio quiet, as a radio-loud object is usually defined to have $R\simgt10$ 
\citep[][]{kellermann89}. 
However, it should be noted that our estimate on the radio loudness by the extrapolation  
is uncertain, as it depends strongly on the radio spectral index which is unknown. 
Recent studies have shown that the radio spectral indices 
are in the range $\alpha\sim$0.5-0.9 \citep[e.g.,][]{greene06, wrobel06, nyland12, reines12, gultekin14}. 
%for low-mass AGNs which have been observed with multiple radio frequencies. 
Assuming a radio spectral index of $\alpha=0.5$ for \src will yield a factor of $\sim$1.5 increase 
in the value of radio loudness.  
%with intermediate-mass BHs in which radio observations at      
%more than one frequency are available 
The $R$ value for J1302 is slightly higher than the majority of the low-mass AGNs studied by 
\citet{greene06}, which have upper limits in the range $R<0.68-9.9$, 
but comparable to the measurements in the extended sample of AGNs with low-mass BHs \citep[][]{greene07b, gultekin14}, 
$0.5<R<60$ with a median value of $\sim7$. 
Note that only eleven of the objects are detected in the FIRST survey ($\sim$6\%, Greene \& Ho 2007b), 
suggesting that low-mass AGNs may have a low incidence of radio activity.    
%Only one object, GH10, was significantly detected in the radio among nineteen sources they studied. 
%The radio detection in J1302 thus makes it unusual when compared to other low-mass AGNs.  
}

\subsection{Infrared observations}
J1302 was detected at 24\um, as 
part of $Spitzer$/MIPS survey program of Coma cluster.  
The 80 percent completeness limit of the survey is $\sim$0.33 mJy.
%, corresponding to a SF rate (SFR) of 0.02 $M_{\odot}$ yr$^{-1}$ at the redshift of Coma \citep{bai06}.
By using a 21\arcsec aperture photometry (corrected by a factor of 1.29 for total), 
the 24\um~flux for J1302 was estimated to be 1.05 mJy \citep{mahajan10}.
{This corresponds to a star formation rate (SFR) of 0.064 \sfr~{\citep[][]{elbaz10}}, if assuming that 
the AGN has little contribution to the 24\ums flux. 
}

J1302 is also included in the {\it Wide-field Infrared Survey Explorer} ($WISE$; Wright et al. 2010) source catalog. 
$WISE$ has mapped the whole sky in 4 bands centered at 3.4, 4.6, 12, and 22\um~(W1, W2, W3, W4), 
 with an angular resolution of 6.\arcsec1, 6.\arcsec4, 6.\arcsec5, and 12.\arcsec0. 
J1302 is detected with a high S/N in W1 and W2, while it is marginally detected 
in W3 and not detected in W4. 
The average magnitude over the set of observations is 
{W1=12.58 mag (2.85 mJy), W2=12.558 mag (1.62 mJy), W3=11.295 mag (0.88 mJy), and W4$<8.6$ mag (2.89 mJy).  
The upper limit on the W4 flux is consistent with the
observed 24\ums flux with $Spitzer$/MIPS. }

To construct the broad-band SED of J1302, we also took near-infrared magnitudes from the 2MASS Point Source Catalog (Skrutskie et al. 2006) in the J (13.31 mag), H (12.67 mag), and $K_s$ (12.336 mag) bands. 
{The corresponding flux density is 7.55 mJy, 8.73 mJy and 7.76 mJy, respectively. }
Given the poor resolution, the 2MASS magnitudes include flux from both the AGN and the host galaxy. 

\section{Results and Discussion}
\begin{figure}
\centering{
\includegraphics[scale=0.8, angle=0]{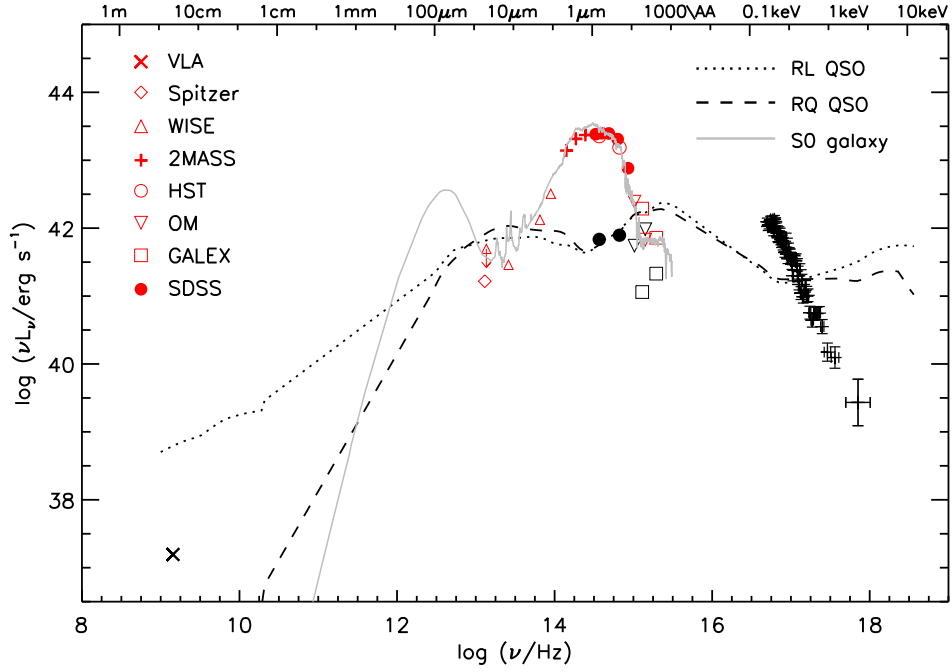}
\caption{
X-ray to radio SED of \src. The AGN emission is plotted in black dots: 
radio data from VLA, optical data from HST, UV data from GALEX and \xmms/OM 
(after the image decompositions),  
%(error bars denote the range of magnitudes allowed by the different GALFIT fitting schemes),
and \xmm X-ray spectrum in the quiescent state. 
% XMM-Newton (1σ uncertainty range given by the dot-dashed line). 
Integrated measurements for the entire host galaxy are plotted with various red symbols. 
We overplot the median SED of radio-loud (black dotted line) and radio-quiet (black dashed line) quasars
(Elvis et al. 1994), scaled to the nuclear optical emission in the HST/F435W band. 
We also plot the SED of a typical S0 galaxy (grey line) from the SWIRE template library of Polletta et al.
(2007), scaled to the HST/F435W optical point for the host galaxy.
%Radio maps from VLA Coma cluster survey (left, Miller et al. 2009) and FIRST.
%The $5\sigma$ detection limit (including clean bias) of J1302 from FIRST survey
%is consistent with the marginal detection (7$\sigma$) from VLA Coma cluster survey.
%No apparent evidence of radio variability can be seen from the two existing VLA observations.
%The circle has size of 5" radius.
%The vertical dotted lines correspond to the FWHM of H$\beta$ line (Wang et al. 2007).}
}\label{fig:sed}}
\end{figure}

We have combined the photometric data described in the previous
section to construct the broadband SED for J1302.  
This is one of the most complete SEDs available for low-mass AGNs 
 {\citep[e.g.,][]{moran99, moran05, thornton08, jiang13}}. The result, plotted in $\nu L_{\nu}$ units
and corrected for Galactic absorption, is displayed in
Figure~\ref{fig:sed}. 
The SED template from a S0 galaxy is overplotted for comparison, as 
it is consistent with the classification of \src from ground-based
imaging. 
Thanks to the high resolution \hst data obtained, we can reveal
partially the nuclear SED of \srcs. 
As Figure~\ref{fig:sed} shows, the optical nuclear SED is at least
ten times weaker than that of host galaxy. 
A comparison with the median SEDs of radio-quiet and radio-loud
quasars (Elvis et al. 1994) suggests that the SED of \src differs
dramatically from  the ones of quasars, particularly from the UV to
the X-ray.  
The obvious big blue bump in optical/UV as seen in typical quasars 
is not observed in \src which is likely shifted to higher frequencies. 
While possible source variability is present in the UV, the \xmms/OM 
measurements appear to agree roughly with disk blackbody model which 
can well fit the X-ray data (S13).  
Further characterization of the optical-to-X-ray data will be presented in
the next Section. 
\subsection{X-ray Spectral Properties}
%Discuss the results from optagn model. 
%\subsubsection{The Nature of Soft X-ray Emission}

As we found in S13, {the X-ray spectrum of J1302 is extreme steep ($\Gamma>7$ in the quiescent state) 
which is unprecedented among AGNs.} 
Modeling the quiescent state spectrum with a disk blackbody yields an 
effective disk temperature of $kT\sim$40 eV, 
{%much lower than the canonical values of $\sim$0.1-0.2 keV for AGNs, 
which is comparable to the expectation from the standard accretion disk model 
(Yuan et al. 2010; S13). 
The fitted disk temperature is, however, in remarkable contrast to the 
canonical temperatures of $\sim$0.1-0.2 keV found for the soft X-ray excess in AGNs 
which are too high to conform to the disk model prediction 
\citep[e.g.,][]{crummy06, ai11}.}
The dominance of thermal disk emission in the spectrum is by analog 
with BHBs in their high/soft state, 
suggesting that the source has a 
{high Eddington ratio ($\simgt0.1$). 
If explaining the ultrasoft X-ray emission with the Comptonization 
of seed photons from the disk, an even higher Eddington ratio of $\simgt0.3$ will be implied \citep{done07}. }
%The presence of strong Comptonization in the flare state (see below)
In combination with the observed luminosity, the BH mass may be at an order
of $\sim10^6M_{\odot}$, consistent with the estimation from optical {(S13)}.  
The {quiescent state however}, requires an additional steep power-law ($\Gamma\sim4$) 
contributing to $\sim$15\% of the total luminosity,  
%(the photon index is consistent with that of the flare state), 
presumably arising from the Comptonization by transient heated electrons in corona. 
It is possible that the additional power-law component likely increases in 
luminosity during the flare state, resulting in a hardening of the X-ray spectrum. 
%The most natural interpretation for the spectral shape is therefore that it represents the
%high-energy tail of the disk emission. 

%On the other hand, the soft X–ray spectrum of J1302 
%may result from strong Comptonization of seed photons from the disk 
%in a corona with plasma temperature $kT_{e}$ and optical depth $\tau_{e}$. 
%However, the temperature and optical depth of the Comptonizing plasma are strongly coupled
%(both are equally involved in shaping the spectrum) and thus cannot be constrained simultaneously. 
%In our fitting with the CompTT model, 
%we fixed the plasma temperature at 20 keV and obtained constraints on the optical depth. 
%The single Comptonized model yields consistent fitting results with the previous blackbody 
%model for the spectrum at both quiescent and flare state.
%The Compton optical depth is
%$\tau=0.16^{+0.07}_{-0.05}$ and $\tau<0.03$ for the flare and quiescent state, respectively.
%Given the lack of X-ray emission above $\sim2$ keV, the properties
%of the putative X–ray corona are largely unconstrained by the data alone.

Here, we consider simultaneous fits to the X-ray and UV data for the two flux states, using a more 
physical model, namely the {\sc optxagnf} model in XSPEC 
\citep{done12, jin12a}.
It assumes that the gravitational energy released in the disk is emitted as a color corrected
blackbody (BB) down to a coronal radius $r_{\rm cor}$, while within this
radius the available energy is distributed between two
Comptonization components: a soft one via Comptonization
in an optically thick cool corona to account for the observed soft X-ray excess, 
and a hard one in an optically thin hot corona to model the standard 
X-ray power-law emission above 2 keV. 
The key aspect of this model is that the optical/UV luminosity 
constrains the mass accretion rate through the outer disk, provided there is an independent 
estimate of the BH mass. 
The parameters of the {\sc optxagnf} model are the BH mass ($M_{\rm BH}$), Eddington ratio 
($L_{\rm bol}/L_{\rm Edd}$), dimensionless BH spin ($a_{\star}$), coronal radius ($r_{\rm cor}$) 
marking the transition from blackbody emission to a Comptonised spectrum, 
the outer radius of the disk ($R_{\rm out}$ in units of $R_g$), 
electron temperature ($KT_e$) and optical depth ($\tau$) for the soft Comptonisation component, 
fraction of the power within $r_{\rm cor}$ which is emitted in the hard Comptonisation component 
($f_{\rm pl}$), and normalization. 
%, which relates 
%the BH mass, Eddington ratio, and BH spin as model parameters. 
%Although 

\begin{figure}
\centering{
\includegraphics[scale=0.8, angle=0]{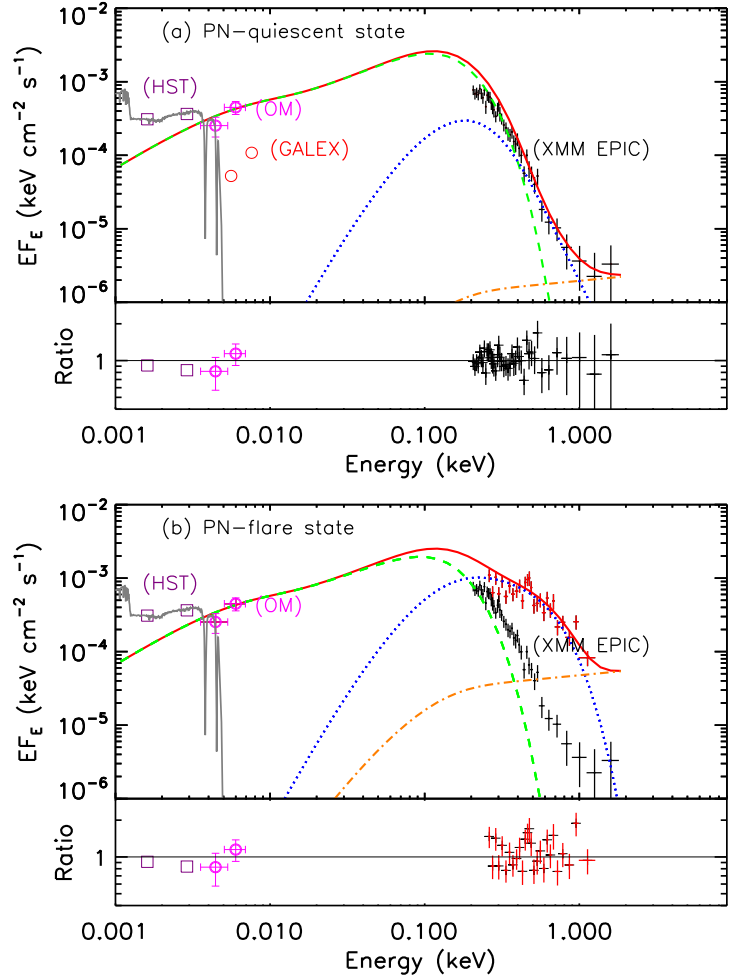}
\caption{
(a) X-ray (quiescent state) to UV SED fitted with the physical model of {\sc optxagnf} (red), consisting of a 
modified accretion disk emission (green dashed), 
a soft X-ray Comptonisation for the soft X-ray excess (blue dotted), and 
a hard X-ray Compotonisation component (orange dot-dashed).   
The GALEX data are not used in the spectral fittings, as they are not taken simultaneously 
and a factor of $\sim9$ lower than the OM ones. 
The \hst~optical data and the corresponding best-fit stellar population model (grey curve) 
are also shown for comparison.
(b) The same {\sc optxagnf} fit but for the X-ray data in the flare state (red crosses). 
The X-ray spectrum in the quiescent state is shown in black for comparison.  
%, likely caused by intrinsic source variability. 
%Joint 99\% confidence contours of the \fekalfa emission line intensity vs. velocity width FWHM, obtained
%from Gaussian fits to the line observed with the \chandra HEG, as described in the text.
%The vertical dotted lines correspond to the FWHM of H$\beta$ line (Wang et al. 2007).}
}\label{fig:xrayuv_sed}}
\end{figure}

However, as we mentioned in Section 2.3.2, the NUV flux from GALEX which is not simultaneous 
with the X-ray data, and is found a factor of $\sim9$ lower than that observed with {\it XMM-Newton}/OM. 
This is likely due to the uncertainty in image decompositions. 
Therefore only the {\it XMM-Newton}/OM data are used in our spectral fits. 
However, there are too many parameters to constrain and 
the UV data are not sufficient to remove the degeneracies, 
% too many parameters to be constrained, 
so we assumed a non–rotating Schwarzschild BH, 
and fixed the outer radius at 10$^5R_g$, the hard powerlaw index at $\Gamma=1.8$.  
%Both Galactic extinction and the small amount of intrinsic absorption from the 
%host galaxy are included in the final model.
{ In our fittings to the quiescent data, we also fixed the coronal radius to 10$R_{g}$ which is the typical value for AGNs \citep{jin12b}, as it is found pegged to the maximum value of 100 $R_{g}$ 
allowed in the {\sc optxagnf} model.  
{In fact, spectral-timing and reverberation studies are suggestive of a 
physically small corona of $<10 R_g$ above the central BH in AGNs 
\citep[e.g.,][]{fabian09, wilkins15}. 
Compact coronae with radii less than $10 R_g$ have also been inferred from X-ray microlensing analyses of some bright quasars \citep[e.g.,][]{chartas09, reis13}.
}
%(e.g., Fabian et al 2009; Wilkins et al. 2015). 
Both Galactic absorption (fixed at $N_H=7.5\times10^{19}$ cm$^{-2}$) and 
intrinsic absorption from the host galaxy are included in the final model.
We accounted for the Galactic extinction by setting 
$E_{\rm B-V}=1.7\times N_{\rm H,Gal}/10^{22}$. 
% as employed in \citep{jin12b}. 
This model provides a good description of the UV and X-ray data with a reduced 
$\chi^2/dof=26.4/40$. The parameters for the {\sc optxagnf} fit are shown in Table 3. 
The best-fit implies a BH mass of $M_{\rm BH}=
2.8\times10^6$\msun~and Eddington ratio $L_{\rm bol}/L_{\rm Edd}$=0.14, which 
are in agreement with those derived from optical spectrum in S13. 
Therefore, if the {\sc optxagnf} model is correct to explain the observed X-ray spectrum for 
\src then a low mass BH accreting at high Eddington rate can be inferred.

Figure~\ref{fig:xrayuv_sed}(a) shows the best-fit {\sc optxagnf} model for the 
X-ray quiescent data (red solid line), 
alongside with different SED components. The thermal emission from accretion disk 
dominates the X-ray spectrum at the quiescent state by accounting for $\sim72$\% of 
the total flux in the 0.2-2 keV, while the Comptonization from low temperature corona 
contributes $\sim27$\%. 
%The high temperature thermal Comptonization component 
%accounts for only $\sim1$\% of the X-ray flux, 
%suggesting that statistically it is not required by the data. 
The best-fit electron temperature and optical depth for the soft Comptonization component is 
$KT_{e}\sim0.1$ keV and $\tau\sim18$, respectively. 
%filter are not sufficient to remove the degeneracies of the Comptonization
%model, and we cannot constrain all the parameters independently. 
%We thus only report one special case in which we assume that the corona is strongly optically
%thick with $\tau_e$ = 13, corresponding 
They are consistent with the average values, i.e., $\langle KT_e \rangle\sim0.3$ and 
$\langle\tau\rangle\sim13$, in the study of a large sample of type 1 AGNs by Jin et al. (2012b), 
suggesting that the Comptonization by an optically thick corona could account for, at least partially, 
the observed soft excess emission.  
The high temperature optically thin Comptonization component, however, 
contributes to only $\sim1$\% of the X-ray flux. 
%suggesting that statistically it is not required by the data. %who used the same {\sc optxagnf} spectral model. 
%This is not surprising as \src is one of few AGNs with the strongest 
%soft X-ray excess discovered so far. 
%Explaining the ultrasoft X-ray excess with the thermal emission from accretion disk
%or its Comptonization 
This suggests that \src may differ from many other AGNs in its environment surrounding the
central engine, i.e., lacking the standard optically thin electron population in corona. 
% and the disk being directly seen in the line of sight. 
%Given the significant detection in the radio (Section 2.4), i
It may be possible that in \src the hot disk corona is strongly suppressed  
due to the formation of jet given its detection in the radio, as found 
in BHBs during their high/soft state.  
However, as the high temperature Comptonization to produce the power 
law emission which dominates above 2 keV is still consistent with the data, 
future longer observations at higher energies than our \xmm data are needed to 
better constrain the coronal parameters.  
We note that, by extrapolating the best-fitting model to the optical emission from the HST 
point source, 
the model under-predicts the flux by a factor of $\sim2$, as shown in Figure~\ref{fig:xrayuv_sed}.
This is likely due to the emission from nuclear star clusters (NSCs), for which we
will discuss in detail in Section 3.2.3.
%However, as no hard X-ray emission above $\sim$ 2 keV is detected, 
%the above estimation on the coronal parameters are still 
%such as the optical depth and the electron temperature can not be made. 
%However, as the X-ray emission above 2 keV is not detected, 
%the coronal physical parameters cannot be constrained
%independently, 
%Future longer observation at higher energies than our \xmm data is needed to
%study to coronal properties by better constraining the electron 
%temperature and optical depth.
%Although our spectral fitting using {\sc optxagnf} model suggesting the extreme condition of 
%corona compared to other AGNs.  
%In this case, 
%From our spectral fitting using {\sc optxagnf}, the extremely soft X-ray emission found in AGNs 
%could be explained by either the thermal emission directly from accretion disk or 
%Comptonized disk emission by corona with unusual physical condition. 
%In either case, the ultrasoft X-ray emission may provide 
%Our spectral
}

{The X-ray data for the flare state along with the best-fit {\sc optxagnf} model are 
shown in Figure~\ref{fig:xrayuv_sed}(b), and the best-fit parameters are summarized in Table 3. 
It was found that when fixing the coronal breaking radius $r_{\rm cor}=10 R_g$, as did for the 
quiescent data, spectral fits with the {\sc optxagnf} model result in an extremely large 
optical depth, which is approaching the maximum value of 
$\tau=100$ allowed in the model. 
The large optical depth is indicative of more low-energy photons being 
upscattered by the Comptonization process.
We then allowed the corona radius to vary in the spectral fits, and the best-fit 
yields a more extended corona of $r_{\rm cor}=21.2R_g$, with a 90\% confidence lower limit at 11.8$R_g$.  
%In the flare state, the optically thick Comptonization component accounts for $\sim$\% 
%of the flux in the 0.2-2 keV, while the thermal emission from accretion disk contributes 
%to $\sim$\%. 
%The increase in the extent of the 
In this situation, 
it is possible that the configuration of the corona is changed before and during the flare, 
e.g., a layer of warm, optically thick 
material over the surface of the accretion disk may arise for a short period. 
Although the physical condition on the corona is unclear, it has been suggested that 
the magnetic reconnection may play a critical role in accelerating the non-thermal electrons 
associated with the corona which could be responsible on the X-ray flares 
\citep[e.g.,][]{wang01, wilkins15, li16}.  
In the flare state, the best-fit of the {\sc optxagnf} model suggests that 
the X-ray spectrum is dominated by the low temperature Comptonization component which 
accounts for $\sim$80\% of the total emission in the 0.2--2 keV.  
%Such a situation has been suggested in magneto-hydrodynamic simulations of  

%Different from the quiescent state, 

%The flare state is dominated by the soft Comptonization component which accounts for $\sim$75\%
%of the flux in the 0.2-2 keV, while the disk thermal emission contributes to only $\sim$16\%. 
%Although we took into account 
%The hard Comptonization component, though included in our spectral fits, contributes little to 
%the observed X-ray emission. 

%We note that, by extrapolating the best-fitting model to the optical data from HST, 
%the model under-predicts the flux by a factor of $>2$, as shown in Figure~\ref{fig:xrayuv_sed}. 
%This is likely due to the emission from nuclear star clusters (NSCs), for which we 
%will discuss in detail in Section 3.2.3.  
} 

%For a given optical depth a steep spectral index means that the electron temperature is low.
%If fixing the optical depth of scattering plasma $\tau=0.1$, we obtained an electron
%temperature of $11\pm2$ keV, much less than the typical temperature of $\sim$100 keV
%found for AGNs (Skibo \& Dermer 1995).
%As the spectra is extremely soft with no hard X-ray emission above $\sim$ 2 keV detected, estimates on
%the coronal parameters, such as the optical depth and the electron temperature can not be made.
%Inclusion of the X-ray data at higher energies than our \xmm data is needed to
%better constrain the plasma temperature and therefore the Compton $y$ parameter.

\subsection{Broadband Continuum Properties}

\subsubsection{The ratio of optical/UV to X-rays}
\begin{figure}
\centering{
\includegraphics[scale=0.7, angle=0]{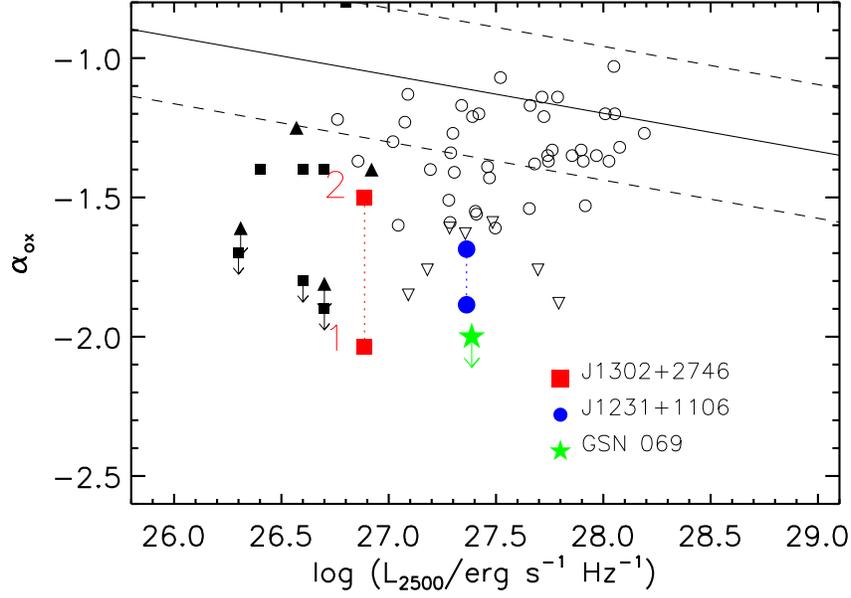}
%\includegraphics[width=0.5\textwidth]{aox.ps}%
%\hfil
%\includegraphics[width=0.5\textwidth]{lo3_lx.ps}%
%\hfil
\caption{
%{\it Left:} 
The optical-to-X-ray spectral index $\alpha_{\rm ox}$ vs. 2500\AA~luminosity for J1302 (red filled squares). 
%The same as Figure??, 
The low-mass AGNs from Dong et al. (2012) are shown in open circles, while the open triangles 
represent upper limits.  
The solid line and dotted lines represent the relation and the 1$\sigma$ scatter given by 
Steffen et al. (2006). 
The symbol ``1" represents low X-ray state for J1302, while ``2" is for high state. 
%The arrow denotes the direction of $\alpha_{\rm ox}$ changing with the AGN 2500\AA~luminosity. 
{For comparison, we plot two ultrasoft AGNs, J1231+1106 (blue filled circles) and GSN 069 (green star), 
and low-mass AGNs from \citet[][]{plotkin16} (filled triangles) and from \citet[][]{yuan14} (filled squares). 
}
%are shown in blue filled circles and green 
%star, respectively. 
%{\it Right:} Correlation between absorption-corrected soft X-ray luminosity in 0.5--2 keV and
%[O{\sc iii}]$\lambda$5007 luminosity for J1302, with two large filled red squares representing
%the flare and quiescent flux values.
%The low-mass AGNs observed with \chandra from the Dong et al. (2012) sample are also plotted for
%comparisons (open circles).
%The solid line is the best-fitting relation from Panessa et al. (2006) with the 68\% scatter
%denoted by the dotted lines.
%The vertical dotted lines correspond to the FWHM of H$\beta$ line (Wang et al. 2007).}
}\label{fig:aox}}
\end{figure}

As we have seen in the above Section, \src appears to have different 
disk-corona geometries from other AGNs. 
In order to gain insights into the unusual X-ray properties of \srcs, we turn 
now to the analysis of the optical-to-X-ray flux ratio. 
Given the extremely weak hard X-ray emission, the ratio in this object
maybe unusually large compared to previously known AGNs. 
The optical-to-X-ray flux ratio is often parameterized by $\alpha_{\rm ox}$, which is defined 
as the effective spectral index between 2500\AA~and 2 keV\footnote{$\alpha_{\rm ox}=\rm log(f_{\rm 2keV}/f_{2500A})/
log(\nu_{\rm 2keV}/\nu_{2500A})$=0.384log($f_{\rm 2keV}/f_{2500A}$)}. 
%It is well known that AGNs are increasingly X-ray weak relative to UV, for higher luminosities.  
%The monochromatic 2500\AA~flux is calculated from the nuclear point source flux 
To calculate the monochromatic 2500\AA~flux, we used the $XMM$/OM flux at 
$\sim2120$\AA~(UVW2)\footnote{Using the monochromatic 2910\AA~flux as an extrapolation results in 
a consistent but slightly flatter $\alpha_{\rm ox}=-$1.95 in the quiescent state.}, % as it is found to be dominated by the AGN, 
assuming an UV spectral index  $\alpha=-0.9~(S\propto\nu^{\alpha})$. 
The $XMM$/OM data which are taken simultaneously with the X-ray observations 
allow for more reliable measurement of $\alpha_{\rm ox}$. 
The X-ray monochromatic flux at 2 keV is estimated from the best-fitting spectral model (S13), 
since it is not detected directly. 
%Note that the estimated X-ray flux is consistent with the $3\sigma$ upper limit. 
In Figure~\ref{fig:aox} we plot the $\alpha_{\rm ox}$ versus $L_{2500\AA}$. 
It is well known that AGNs are increasingly X-ray weak for higher 
UV luminosities.  
Such relation for the optically selected AGNs is shown by the solid line 
and the dotted line is the 1$\sigma$ scatter (Steffen et al. 2006). 
J1302 is displayed as red filled square in the flare and quiescent state, respectively.  
%while other two similarly ultrasoft AGNs, J1231 and GSN069, are shown in ??, respectively.
%and low-mass AGNs with high-Eddington ratios (Dong et al. 2012) are also plotted for comparisons (open circles). 
%For comparison, we also plot low-mass AGNs with high-Eddington ratios (Dong et al. 2012).  
It shows that the $\alpha_{\rm ox}$ for J1302, like some low mass AGNs 
with high-Eddington ratios (Dong et al. 2012), 
falls below the low-luminosity extrapolation of the previously found $\alpha_{\rm ox}-L_{2500\AA}$ 
relation. 
In particular, J1302 in the quiescent state is significantly (hard) X-ray weak compared to typical 
AGNs ($>10$ times weaker than expected with respect to its optical/UV emission). 
{The $\Delta\alpha_{\rm ox}=\alpha_{\rm ox}-\alpha_{\rm ox,exp}=-0.46$ and $-0.99$ for 
the flare and quiescent state (see Table 4)\footnote{$\alpha_{\rm ox,exp}$ is the average $\alpha_{\rm ox}$ 
expected from the Steffen et al. (2006) $\alpha_{\rm ox}-L_{\rm 2500}$ relation.}, 
corresponding to being X-ray weaker than $\alpha_{\rm ox,exp}$ at the 2.15$\sigma$ and 10.64$\sigma$ level, respectively 
\citep[e.g.,][]{plotkin16}.}
%AGNs ($\sim10$ times weaker than expected in the X-ray at 2 keV).
%{\bf However, it should be noted that the UV luminosity which was derived from the GALFIT decomposition 
%may describe an upper limit on the AGN emission (Section 2.3.1). 
%Thus the true $\alpha_{\rm ox}$ for \src maybe slightly higher. }
For comparison, we also plot the $\alpha_{\rm ox}$ for the ultrasoft AGN J1302+2746 and GSN 069 
(Terashima et al. 2012; Miniutti et al. 2013), which are found to be significantly lower than the extrapolation of the $\alpha_{\rm ox}-L_{2500\AA}$ 
relation, indicating similar hard X-ray weakness.  
%display similar extremely soft X-ray emission (Terashima et al. 2012; Miniutti et al. 2013). 

%do not follow the Steffen et al. (2006) relation, and is clearly X-ray weak. 

\begin{figure}
\centering{
\includegraphics[scale=0.8, angle=0]{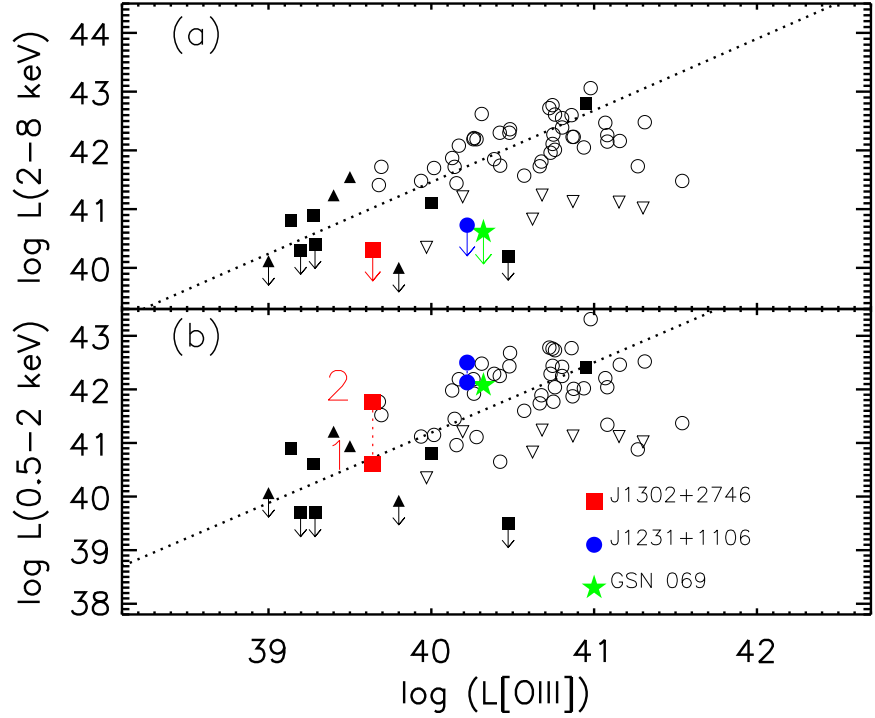}
\caption{
{\it Upper:} Correlation between absorption-corrected hard X-ray luminosity in the 2--8 keV and
[O{\sc iii}]$\lambda$5007 luminosity for J1302.
%, with two large filled red squares representingthe flare and quiescent flux values.
%The low-mass AGNs observed with \chandra~from the Dong et al. (2012) sample (open circles) 
%and Yuan et al. (2014) are also plotted for comparisons (filled triangles).
{Symbols have the same codings as in Figure 6}. 
The solid line is the best-fitting relation from Panessa et al. (2006). 
% with the 68\% scatterdenoted by the dotted lines.
{\it Lower:} The same as above, but for the correlation between soft X-ray luminosity in the 0.5--2 keV and [O{\sc iii}]$\lambda$5007 luminosity. Filled red squares represent 
the \src~flux in the flare and quiescent state respectively.   
%Joint 99\% confidence contours of the \fekalfa emission line intensity vs. velocity width FWHM, obtained
%from Gaussian fits to the line observed with the \chandra HEG, as described in the text.
%The vertical dotted lines correspond to the FWHM of H$\beta$ line (Wang et al. 2007).}
}\label{fig:lo3_lx}}
\end{figure}

%The apparent X-ray weakness of J1302 
%may imply that its coronal condition is unusual. 
%could be simply attributed to 
%the absence of the standard (optically thin) 
%X-ray corona or its inability to efficiently up-scatter the soft disk photons.
A similar perspective comes from comparing the hard X-ray luminosity with 
optical line emission, e.g., {[O\sc iii]}, as the two are strongly correlated 
in unobscured AGNs \citep[e.g.,][]{heckman05, panessa06}. 
%(e.g., Heckman et al. 2015; Panessa et al. 2006). 
The hard X-ray (2--8 keV) and {[O\sc iii]} line luminosity relation is shown in Figure~\ref{fig:lo3_lx}(a). 
We include the best-fitting relation from a sample of Seyfert galaxies with high BH 
masses (e.g., Panessa et al. 2006), and low-mass AGN sample from Dong et al. (2012) for comparison. 
%J1302 and the other two hard
The intrinsic 0.5--2 keV luminosity of J1302 is $L_{0.5-2 \rm keV}
\sim1.2\times10^{42}$ \erg, while the 2--8 keV luminosity is 
$L_{\rm 2-10 keV}\leq2.03\times10^{40}$ \erg. 
J1302, together with the other two ultrasoft AGNs, do lie below the relation 
defined by the higher luminosity sources. 
%Miniutti et al. (2009) presented that the 0.5--2 keV and 2--10 keV 
%luminosities obey a tight correlation in a sample of PG quasars and AGN 
%with low mass BHs. 
According to this correlation the observed upper limit on $L_{\rm 2-8 keV}$ implies that the 
hard X-ray emission in J1302 is at least five times fainter than in typical 
AGNs. 
{Note that given the low luminosity of the \src nucleus, a fraction of 
the {[O\sc iii]} line luminosity may be associated with the star formation process. 
Using the {[O\sc ii]$\lambda$3727} emission line as a tracer of star formation, 
and the observed median {[O\sc iii]/[O\sc ii]} ratio of --0.54 dex for galaxies \citep[][]{moustakas06}, 
we obtained a SFR of $\sim$0.21 \sfr assuming the {[O\sc iii]} line is due to star formation activity. 
This SFR is a factor of three larger than that obtained from the IR luminosity (Section 2.5), 
suggesting that most, if not all, of the {[O\sc iii]} is ionized by AGN emission. 
This is supported by the observed ratios of the narrow lines {[O\sc iii]/$\rm H\beta>4.8$} and 
{[N\sc ii]/$\rm H\alpha=2.3$}, which place \src into the Seyfert regime on 
the BPT diagram of \citet[][]{kewley06}.    
Subtracting the SFR contribution from the {[O\sc iii]} luminosity, 
the hard X-ray emission for \src is still 
three times fainter than typical AGNs. 
 }
%This translates into a 2--10 keV X-ray Bolometric correction 
%$k_{2-10 \rm keV}\geq$??, which is one order of magnitude lower than the 
%typical value for AGNs obtained in Vasudevan et al. (2009). 

We argue that the X-ray weakness of J1302 is not due to intrinsic
absorption, though it has been proposed to play a role in some IMBH AGNs
(e.g., Dong et al. 2012; Yuan et al. 2014). 
J1302 displays a highly variable soft X-ray emission (S13),  and its 
X-ray spectra show no evidence of any significant intrinsic X-ray
absorption. 
Comparing the 0.5--2 keV emission with the strength of the {[O\sc iii]} line further supports this, as illustrated in
Figure~\ref{fig:lo3_lx}(b). 
%We argue that the lack of any significant intrinsic absorption in its X-ray spectrum 
%The highly variable soft X-ray emission and lack of any significant intrinsic absorption in its X-ray spectrum 
%strongly support this. 
%This can be examined by comparing the X-ray luminosity with the [\sc oiii]. 
%The correlation between 0.5-2 keV luminosity with [\sc oiii]$\lambda$5007 luminosity is shown in Figure??. 
%In Figure~\ref{fig:lo3_lx}, we plot the correlation between 0.5-2 keV luminosity with {[\sc oiii]}$\lambda$5007 luminosity. 
The ratio of soft X-ray to {[O\sc iii]} luminosity can be used as an indicator of X-ray absorption.  
Obscured AGNs fall off the correlation being underluminous in X-rays for 
a given {[O\sc iii]} luminosity.   
%We include in Figure~\ref{fig:lo3_lx} the best-fitting relation from a sample of Seyfert galaxies with high BH 
%masses for comparison (e.g., Panessa et al. 2006), and low-mass AGN sample from Dong et al. (2012). 
%For consistency with the J1302 measurement, X-ray luminosities between 0.5--2 keV are used 
%for all literature sources.  
As can be seen,  
%especially at the quiescent state, appears fainter than other 
%unobsorbed low-mass type 1 AGNs as well as the two similarly ultrasoft IMBHs, 
%and extend  
%the ratio between X-ray and {[\sc oiii]} luminosity for i
J1302 do lie within the $L_{\rm X}$--$L_{\rm [O~III]}$ relation defined by 
unabsorbed type 1 AGNs, both of high and low BH mass. 
%In contrast, the four low-mass type 2 systems studied by Thornton et al. (2009) all
%sit substantially below the bulk of the type 1 AGNs in Figure~\ref{fig:lo3_lx},
%consistent with the X-rays being significantly absorbed in these
%systems. 
% and extend the relation to lower luminosity regime, 
%but it is much fainter compared to the other 
%two extremely soft IMBHs.  
Another trend can be seen is that J1302, especially at its quiescent state, 
%extend the $L_{\rm X}$--$L_{[OIII]}$ relation down to fainter
%than their high-accretion counterparts, and 
extends the $L_{\rm X}$--$L_{\rm [O~III]}$ relation down to the lower luminosity regime 
for low mass AGNs (Yuan et al. 2014). 
%The trend is roughly consistent with
%the extrapolation of the previously known relations for 
%AGNs with high BH masses (e.g., Panessa et al. 2006), albeit with large scatters.

%at flare state appears to be consistent with the 
%Steffen et al. relation,    
%\subsection{On the Lack of Broad Emission Lines}
%The lack of evidence of X-ray absorption for J1302, which is also supported by 
%the highly variable soft X-ray emission (S13), is suggestive of
%The above analysis suggests the ex X-ray luminosity for J13. 
%that is comparable to that of the low $L_{\rm bol}/L_{\rm Edd}$ IMBHs (Yuan et al. 2014). 
%In this respect the object 
%Therefore, 
%The lack of evidence of X-ray absorption is suggestive of intrinsically X-ray weakness of J1302,
The above analysis strongly suggests that J1302 is intrinsically X-ray weak, 
though the true nature of which is unclear.  
%the nature of which is unknown. 
One possibility is that J1302 may represent an extreme case of a
low BH mass but high Eddington ratio system in which the standard 
hard X-ray corona is absent or unable to efficiently up-scatter the disk photons 
{\citep[e.g.,][]{leighly07, miniutti13}}. 
%This may explain the apparent X-ray weakness of J1302. 
%Indeed, as shown in our spectral modelling, it is quite striking that 
%the UV-to-X-ray data are dominated by an accretion 
%disk spectrum with a relatively weak X-ray coronal contribution. 
%The disappearance of corona could naturally explain 
%This is likely due to the very large Eddington ratio in \srcs. 
%In this 
%Among the possible explanations for the lack/weakness of optically thin coronal emission,
%we point out the work by Proga (2005) who discusses how
%One specific mechanism dd
When the Eddington ratio is high ($L_{\rm Bol}/L_{\rm Edd}\simgt0.3$), 
the structure of the standard thin accretion disk changes to a geometrically and 
optically thick slim accretion disk, where outflow and/or disk-wind becomes important. 
In this situation, a dense, highly ionized ``failed wind" produced by overionization 
could fall into the corona and thus may quench the hard X-ray emission through 
bremsstrahlung (Proga 2005).  
%%a failed disk-wind may quench the coronal X-ray emission. In
%that situation, the relatively high density in the failed wind means
%that bremsstrahlung losses dominate over inverse Compton, thus
%quenching the hard X-ray emission, while preserving the soft X-rays.
%??Explain why X-ray weak.?? 

%ADD BAL quasar and slim disk scenario. 

%%Dong et al. (2012) have suggested that the X-ray weakness in their IMBHs sample is 
%likely casued by 

\subsubsection{Intrinsically lack of BLR?}

\begin{figure}
\centering{
\includegraphics[scale=0.9, angle=0]{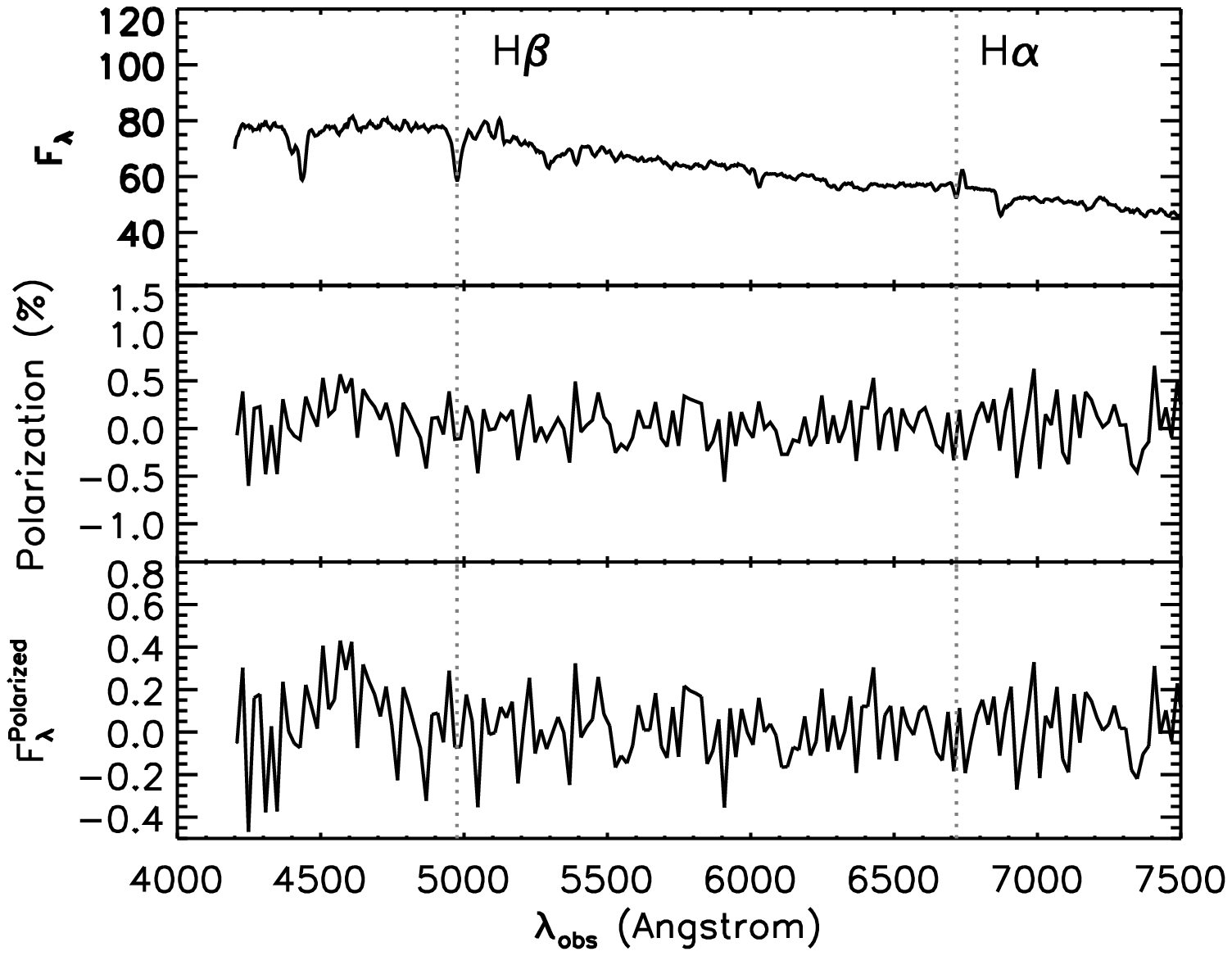}
\caption{
Spectropolarimetry of \src. Top: total flux spectrum ($F_{\lambda}$); middle: polarization 
fraction; bottom: polarized flux spectrum. The flux scales are in units of 
$10^{-17}$\ergs$\AA^{-1}$.  
The position of H$\alpha$ and H$\beta$ is marked (vertical dotted line). 
The polarization and polarized flux spectrum are binned to have a size of 20\AA. 
The polarization is measured along the direction with largest signal.  
Although a small polarization (including instrumental one) is detected, 
no polarized broad lines are seen in the polarized flux spectrum. 
}\label{fig:mmt}}
\end{figure}

Note that, like the other two ultrasoft X-ray AGNs, J1302 is 
unabsorbed in the X-rays but do not show any broad Balmer lines in the optical, suggesting
that this AGN may lack the BLR or that the corresponding emission
lines are much weaker than in typical AGN, as suggested by Miniutti et al. (2013).
This is puzzling and in apparent contradiction with the unification model of AGNs 
(Antonucci 1993). 
{The contradiction might be attributed to a patchy torus, which could block 
most of the BLR emission while enable occasional leakage of the central X-ray 
emission. If this is the case for J1302, we would expect X-ray spectral transitions, 
i.e., from unabsorbed to heavily obscured (to be testified with future X-ray
observations). Meanwhile, we also expect a hidden BLR.} 
We have therefore conducted a pilot search for weak or hidden broad emission lines (BELs) in \src 
using optical spectropolarimetry. 
The results are shown in Figure~\ref{fig:mmt}. 
Although a small polarization (including instrumental one) is detected, no polarized broad H$\alpha$ or H$\beta$ is seen, 
suggesting that \src may intrinsically lack a BLR. 
{However, the spectral resolution of our observation is not high, 
which may not be efficient to detect broad $H\alpha$ in polarization if its FWHM is less than 700 km/s. 
In addition, } as the object was observed in a relatively short (40 min) exposure and 
the S/N is poor in the spectropolarimetry data,  
deeper spectropolarimetric observations are needed to firmly 
exclude the presence of weak or hidden BELs. 
%deep search for any weak or hidden BLR is required. 
%Moreover, no narrow emission lines
%are visible in the polarized flux spectra, indicating that any polarization
%imposed outside of the host galaxies is negligible. 

%The slit was aligned with atmospheric dispersion and the rotator was set to track field rotation. 
Models have suggested that if the BLR is part of an outflow, or disk-wind, it 
is unable to form once the AGN accretion rates below the critical value of 
$\sim10^{-3}$ \citep{nicastro00, elitzur09}. 
%(Nicastro et al. 2000; Elitzur \& Ho 2009). 
In this scenario, 
J1302 appears to be a puzzling ``true" Seyfert 2 candidate with Eddington ratio 
much higher than the critical value for the BLR disappearance. 
\citet{miniutti13} proposed that the lack of BLR in GSN 069, another ultrasoft 
AGN similar to \srcs, maybe attributed to either the lack of hard X-ray emission 
in a two-phase BLR model, or an evolutionary scenario in which the BLR has not fully formed. 
Such idea can be tested in \src by detecting and following the evolution of the hard 
X-ray emission (if present). The (quasi)-simultaneous optical/UV spectroscopic campaign 
may be useful to confirm/dismiss the above interpretation. 

%We now examine 
%We computed a few simple photoionization models
%using the code CLOUDY (Ferland et al. 1998) to test whether
%the unusual X-ray spectrum of J1302 can be responsible
%for the excitation of its optical lines. We assume that the
%ionizing continuum can be described by a blackbody with
%$T = 1.7 \times 10^6$ K, corresponding to a ?? eV blackbody that
%can approximate the X-ray spectrum.
%??TBD???These calculations suggest that the observed
%X-ray spectral energy distribution of J1302 can, indeed,
%reproduce line ratios that span the locus of AGNs??.

%\subsubsection{The Presence of Nuclear Star Cluster?}

\subsection{Host Galaxy Structure and BH-Host Connection}

From our GALFIT decompositions, we conclude that the
host galaxy can be well fitted in the both HST filters by a
\sersic\ profile with $n\sim1.5$ and $r_{e}\sim$0.\arcsec41 at $r<$2\arcsec, 
and an exponential profile (equivalent to $n=1$ \sersic\ function that is usually used 
to describe disk at the outer region, see Figure 1).  
The structure of the inner region of the host galaxy remains complicated due
to the slight excess of emission between $\sim0.2-0.5$ arcsec in the F435W filter, which 
could be attributed to the PSF mismatch. 
%second Se´rsic component needed to provide a reasonable
%fit in the F435W filter. We
Looking at the inner region, there is a significant vertical ``X"-shaped structure
in the residual image (Figure~1).
Such a structure has also been reported by \citet[][]{caldwell99}, 
and not unusual for Milky Way mass galaxies
in the local Universe \citep{laurikainen14}, which is confirmed
by N-body simulation as a natural evolution result from a pure disk galaxy (Li \& Shen 2012). 
By excluding the AGN emission, we find that the inner \sersic\ component contributes 
$\sim15\%$ of the total flux of the host galaxy. 
Such the ratio between the bulge and total light ($B/T$) is consistent with 
the median value ($B/T=0.16$) found for the galaxies hosting low-mass BHs \citep{jiang11a}. 
The flatter index ($n<2$) and small B/T suggest the bulge of \src to be pseudobulge. 
As the pseudobulges are believed to be formed by secular processes driven by disk instabilities, 
including the slow rearrangement of material by bars, oval disks, and spiral structure, the 
presence of a low-mass AGN in J1302 seems to support the hypothesis that low-mass BHs 
evolve secularly and are most likely not fueled by major mergers \citep{kormendy04}. 

%Considering
For galaxies with classical bulges, BH masses have been found to correlate with bulge luminosity 
(the $M_{\rm BH}-L_{\rm bulge}$ relation, e.g., Marconi \& Hunt 2003). 
%Considering their different formation mechanisms of classical bulges and pseudobulges, 
Considering that pseudobulges have quite different properties from those rapidly formed 
classical bulges, we may expect differing BH scaling relations as well, which are observed 
with small samples with dynamical BH masses \citep{greene08}. 
We discuss shortly the connection between the BH in \src and the bulge luminosity of 
its host galaxy. The lack of detectable BELs from either direct and polarized optical lights  
prevents us from estimating the mass of the central BH using conventional virial method. 
If we employ the $M_{\rm BH}-\sigma_{\star}$ relation at the low-mass end by Xiao et al. (2011) 
and use the velocity dispersion traced by narrow lines,  
we find a BH mass of $M_{\rm BH}=8\times10^5$\msun ~for \src (S13). 
For the estimate an absolute magnitude for the bulge, we 
use the F435W B-band magnitude of the corresponding \sersic\ component, 
which yields $M_B\sim-17.34$ for the host galaxy. 
Extrapolating the \citet{graham07} $M_{\rm BH}-L_{\rm bulge}$ relation down to a lower
luminosity, we find a predicted mass of $M_{\rm BH}=2.5\times10^7$\msun, 
which is a factor of $>30$ higher than current estimate. 
{Adopting the recently revised $M_{\rm BH}-L_{\rm bulge}$ relation by \citet[][]{mcconnell13}, 
we obtain a predicted mass of $M_{\rm BH}=1.8\times10^7$\msun (assuming a conversion factor of $B-V$=0.9 for the 
bulge lights, Bentz et al. 2009).}
Using the $I$-band absolute magnitude and the \citet{bentz09} $M_{\rm BH}-L_{\rm bulge}$ 
relation result in a similar higher BH mass of $M_{\rm BH}=2.2\times10^7$\msun.  
Note that the pseudobulge in \src may contain younger stars which would affect 
the estimate on the $B$-band bulge luminosity. 
We find $B-I=1.33$ mag for the host galaxy, which is close to the typical color of 
S0/Sab galaxies found in Fukugita et al. (1995), $B-I\sim1.8$ mag, suggesting 
that the correction for age or mass-to-light ratio to the bulge luminosity is small.  
Therefore, though with large uncertainty associated with the BH mass measurement,  
\src, like other pseudobulges containing low-mass BHs (Greene et al. 2008; 
Jiang et al. 2011a), appears to deviate from the $M_{\rm BH}-L_{\rm bulge}$ relation of
classical bulges and elliptical galaxies. 
%Bear in mind that the absolute positions of our objects in the MBH–Lbulge plane are uncertain
%due to potential systematic offsets in the value of f and
%other systematic uncertainties associated with the BH mass
%measurements, as well as the unknown mass-to-light ratios of
%the bulges.
 
%We will discuss how each of these uncertainties may
%impact our conclusions. 

%\subsection{Comparison with???}
\subsubsection{Nature of the Extra Lights in the Optical}
%We found in Section 3.1 that while the {\it XMM-Newton} UV and the X-ray data can be well described 
%by thermal emission from an accretion disk, there is a clear excess of the 
%flux mostly in the optical compared to the best-fitting model. 
As we mentioned in Section 3.1, fitting the {\it XMM-Newton} UV and the X-ray SED with 
{\sc optxagnf} model identified the presence of a significant optical excess in the \hst~bands, 
although this model accounts for the X-ray emission very nicely. 
%{\bf The excess is even more significant if we considering the derived UV emission for AGN as 
%an upper limit (Section 2.3.1)}. 
The unresolved morphology from the \hst~suggests that the optical excess emission 
comes from a very compact region of $r<47$ pc, which is likely associated with a NSC. 
%This extra optical light likely comes from the stellar emission such as NSC.  
The coexistence of NSCs with BHs is not unusual and has been inferred for many low mass AGNs
\citep[e.g.,][]{seth08, seth10}. 
%(e.g., Seth et al. 2008; 2010).
For example, it has already been demonstrated that the first two prototype of this kind,
NGC 4395 and POX 52, has both an AGN and a nuclear star cluster in its center 
\citep{filippenko03, thornton08}.  
%(Filippenko \& Ho 2003; Thornton et al. 2008).

We attempted to fit the optical excess in the SED (\xmms/UVW1, \hst/F435W and \hst/F814W) using stellar population models 
representing emission from a stellar cluster. 
We used the Maraston (2005) simple stellar population models which are constructed 
assuming the Salpeter initial mass function. 
These models provide spectra over a finely spaced age and metallicity grid\footnote{See http://www.maraston.eu}. 
%and at metallicities of Z=0.0001, 0.0004,
%0.004, 0.008, 0.02, and 0.05 at a spectral resolution similar to
%our observations. 
However, the metallicity ($Z_{\star}$) of the stellar population could not
be constrained and so we froze it to Solar values as it is the approximate midpoint of the
metallicity range in the Maraston (2005) model.  
The best fit model is shown in Figure~\ref{fig:xrayuv_sed} (grey curve). 
The age of the stellar population was found to be $\sim$80 Myr.   
Changing the metallicity to lower values results in higher ages. 
For instance, fixing metallicity at $Z_{\star}=0.02Z_{\sun}$ and $Z_{\star}=2Z_{\sun}$ 
yields a stellar age equal to 300 Myr and 65 Myr, respectively.   
%change the other parameters of 
%the results significantly, indicating that the fit is completely insensitive to metallicity. 
Note that previous studies have found significant evidence for multiple
stellar populations in NSCs (e.g., Siegel et al. 2007; Seth et al. 2010), 
but the statistics of our data are insufficient to allow us to attempt to fit 
for an additional stellar contribution. 

Although the formation mechanisms for NSCs are still under debates, they may 
lead to the formation of central BHs. 
% in hierarchical galaxy formation scenarios. 
One possibility is that NSCs are created from gas accretion onto the 
nucleus due to disk gas dynamics 
%galaxy merging (Mihos \& Hernquist 1994) or from disk gas dynamics 
(Bekki et al. 2006). This in situ scenario is favored by observations
that NSCs in spiral galaxies have complicated star formation histories, 
suggesting frequent episodic star formation (Walcher et al. 2006; Rossa et al. 2006). 
Indeed, Caldwell \& Rose (1997) have shown that \src is a very strong post-starburst galaxy with 
a post-starburst age of 500 Myr, based on the presence of enhanced Balmer absorption lines 
compared to those in normal galaxies (see also S13). 
It is quite possible that the formation of central BH in \src was connected with the 
starburst event. 
The presence of a young stellar population with ages up to $\sim300$ Myr around the \src nucleus 
seems to reconcile this picture, 
%with NSC in \src residing in a starburst, 
indicative of further gas accretion onto the nucleus. 
If our interpretation of optical excess component with NSC is correct, we could be expecting 
its mass is the same order as BH, i.e., $\sim1\times10^6$\msun. 
Future high-resolution observations with Integral Field Unit spectroscopy will be helpful to 
construct a dynamical model to estimate the mass (and mass-to-light ratio) 
of the NSC and of a central BH inside it.  
%optical spectra where the enhanced Balmer absorption lines have 
%been observed. 

%It should be noted,
%however, that the large uncertainties in the luminosity (in
%part due to the large errors in the absorption and extinction),
%age, and metallicity of the stellar population produce
%very large errors in the stellar mass. 

\section{Conclusions}

We have conducted follow-up multiwavelength study of the nuclear and host galaxy properties 
of \src, a newly discovered low mass galaxy that displays extremely soft and variable 
X-ray emission. 
%Taking use of the \xmm 
We have shown that the UV-to-X-ray SED can be well and self-consistently described
by thermal emission from an optically thick accretion disk around a BH, with a 
optically thin Comptonized emission from corona. 
The derived parameters ($M_{\rm BH}$ and $\dot{m}$) 
from modeling agree with the independent estimates based on the optical data. 
We thus consider the ultrasoft X-ray emission could be 
a signature of X-rays from an accretion disk around an IMBH. 
%possible evidence for BH accretion disks in AGNs. 

The source appears intrinsically X-ray weak, especially in the quiescent state, 
with hard X-ray emission at least three times fainter than in typical AGNs. 
The lack of hard X-ray emission may be connected with the absence of 
broad optical lines in \srcs.  
We performed a pilot search for weak or hidden broad emission lines 
using optical spectropolarimetry observations. Although a small polarization is 
detected, no polarized broad $H\alpha$ (or $H\beta$) is seen, indicating that \src  
likely intrinsically lacks a BLR. 
%Unexpectively, 
\src is detected significantly in the radio, which is most likely related to 
the accreting low-mass BH. But the radio emission is not strong enough to qualify the 
source as radio loud.  
%As such, \src is an AGN analog of BHXBs at the high/
%soft state in a sense that the accretion disk emission dominates
%the bolometric luminosity. 

We performed a comprehensive analysis of the \hst~images in $B$ and $I$ bands 
to quantify the structure and morphology of the host galaxy. 
The galaxy is an isolated, nearly edge-on S0 galaxy with a relatively weak 
pseudobulge (\sersic~index $n=1.45$) that accounts for $\sim$15\% of the total light of the host galaxy. 
The inner structure of the galaxy is complex, with a significant ``X"-shaped structure in 
the residual image, which may result from the internal disk dynamical instabilities. 
% in the $B$ band. 
The pseudobulge has relatively blue colors ($B-I\sim$1.33 mag). 
Consistent with other low-mass AGNs, \src seems to deviate from the 
$M_{BH}-L_{bulge}$ relation of classical bulges and elliptical galaxies. 
Although unresolved in the morphology, we identified a significant optical excess 
in the \hst~ bands from the SED modeling, which is likely due to the presence of 
a NSC. %The inferred age of the stellar population 
%Although the age of the stellar population cannot be uniquely constrained, it appears that 
%the formation of NSC as well as the central BH in \src is connected with the post-starburst event, 
%in which the gas accretion onto to nucleus due to disk instabilities plays a major role. 
%Further high resolution observations with Integral Field Unit spectroscopy are required to confirm this scenario.  
{Further high resolution UV observations with \hst~are required to confirm this scenario, 
and helpful to further constrain the stellar properties of NSC if present.   }

%there is a clear optical excess in the 
%\hst~ bands compared to the best-fitted SED model, 
%The age of the population cannot be uniquely constrained, with both young and old stellar populations allowed. 

%highly suggestive of a dominant AGN despite having no optical signatures of accretion
%activity. Our main
%Figure 1 shows an example of the \herschel selected dust obscured galaxy at $z=2.9$ for which
%the \herschel SPIRE flux densities are probably mis-matched to a closer and brighter $z=0.9$ source in
%the 24$\mu$m. This is confirmed by the mid- to far-infrared SED fitting to the deblended photometry, which
%is consistent with a $z=2.9$ starburst, permitting to perform a "spectral deconfusion". 
%The spectral doconfusion technique allows to deconvolve the $500\mu m$ flux 
%down to $\sim$5 mJy level for the $Herschel$-selected sources, 
%comparable to the confusion limit in this band (Nguyen et al. 2010; Elbaz et al. 2011). %thus probe fainter galaxy populations than other 

\begin{acknowledgements}
This work was supported by Chinese NSF through grant 11573001, 11233002, 11421303,  
and National Basic Research Program 2015CB857005.
X.S. acknowledges support from the Fundamental Research Funds for the Central Universities (WK3440000001), Anhui Provincial Natural Science Foundation (1608085QA06), and 
the Open Research Program of Key Laboratory of Space Astronomy and Technology, CAS (KLSAT201601). 
J.W. acknowledges support from the CAS Frontier Science Key Research Program (QYZDJ-SSWSLH006).  
This work has made use
of the data obtained through % of the SDSS and data obtained through Lijiang 2.4m telescope and 
the Telescope Access Program (TAP) in 2016A.  
%Partial support for this work was provided by 
 
%work was provided by National Basic Research Program
%2015CB857004..

\end{acknowledgements}

%\clearpage
%\input{photometry}

\clearpage
%\input{table}
%\clearpage

\clearpage

%\begin{figure}
%\centering{
%\includegraphics[scale=0.9, angle=0]{../figure/lo3_lx.ps}
%\caption{
%Correlation between absorption-corrected soft X-ray luminosity in 0.5--2 keV and 
%[O{\sc iii}]$\lambda$5007 luminosity for J1302, with two large filled red squares representing 
%the flare and quiescent flux values. 
%The low-mass AGNs observed with \chandra from the Dong et al. (2012) sample are also plotted for 
%comparisons (open circles). 
%The solid line is hte best-fitting relation from Panessa et al. (2006) with the 68\% scatter 
%denoted by the dotted lines. 
%Joint 99\% confidence contours of the \fekalfa emission line intensity vs. velocity width FWHM, obtained
%from Gaussian fits to the line observed with the \chandra HEG, as described in the text.
%The vertical dotted lines correspond to the FWHM of H$\beta$ line (Wang et al. 2007).}
%}\label{fig:lo3_lx}}
%\end{figure}

%\newpage
%\clearpage

%\clearpage
%\begin{figure}
%\centering{
%\includegraphics[scale=0.9, angle=0]{figure/m0429_radec.ps}
%\includegraphics[scale=0.9, angle=0]{m0429_hdrop_img.ps}
%\caption{
%Joint 99\% confidence contours of the \fekalfa emission line intensity vs. velocity width FWHM, obtained
%from Gaussian fits to the line observed with the \chandra HEG, as described in the text.
%The vertical dotted lines correspond to the FWHM of H$\beta$ line (Wang et al. 2007).}
%}}\label{fig:redshift}

\clearpage

\begin{deluxetable}{c|ccclcccc}
%\tiny
%\tabletypesize{\small}
\tablewidth{0pt}
\tablecaption{GALFIT Decomposition of \hst\ Images\label{tbl-galfit}}

%\tablenum{2 . GALFIT Decomposition}
\tablehead{ \colhead{Filter} & \colhead{Component} & \colhead{$m$} & \colhead{$M$} & \colhead{$f_{\lambda}$}
& \colhead{$n$} & \colhead{$r$ (\arcsec/pc)} & \colhead{$b/a$} & \colhead{$c$} \\
(1) & (2)    & (3)  & (4) & (5) & (6) & (7) & (8) & (9)}
\startdata
F814W  & PSF      & 18.12$\pm0.14$ & $-16.81$$\pm0.14$ &  $0.64\pm0.09$   &\nodata   & \nodata    & \nodata & \nodata   \\
       & \sersic\ & 16.39$\pm0.34$ & $-18.67$$\pm0.34$  & $3.16\pm1.0$   &1.45  & 0.41/193      & 0.54    & 2.68      \\
       & Exp Disk & 14.50$\pm0.17$ & $-20.56$$\pm0.17$  & $18.0\pm2.82$   &[1.0] & 2.86/1332     & 0.24  & \nodata     \\
F450W  & PSF      & 19.04$\pm0.14$ & $-16.06$$\pm0.14$  & $1.37\pm0.18$   &\nodata &  \nodata    & \nodata   & \nodata \\
       & \sersic\ & 17.57$\pm0.34$ & $-17.34$$\pm0.34$  & $5.29\pm0.68$   &[1.45]  & 0.37/178    & 0.52 &  1.49        \\
       & Exp Disk & 16.05$\pm0.17$ & $-18.89$$\pm0.17$  & $21.4\pm2.76$2   &[1.0]   & 2.69/1280   & 0.24 & \nodata      
\enddata
\tablecomments{ Col. (1): $HST$ filter.
Col. (2): Components used in the fitting schemes.
Col. (3): The integrated magnitudes on the Vega system, not corrected for
Galactic extinction. 
Col. (4): The absolute Johnson $I$ and $B$ magnitude after Galactic extinction 
correction. We assume a power-law continuum $f_\nu \propto \nu^{-0.5}$
for the central AGN, an Sb galaxy for the \sersic\ component, 
and an $S0$ galaxy for the disk, using templates from Kinney et al (1996).
{Col. (5): Flux density in units of $10^{-16}$\ergs$\AA^{-1}$.
}
Col. (6): The \sersic\ index.
Col. (7): The effective radius of the \sersic\ component or scale length of
exponential disk, in units of arcsec and pc, respectively.
Col. (8): Axis ratio.
Col. (9): Diskness (negative)/boxiness (positive) parameter,
defined in Eqn.~(3) of Peng et al. (2002).
The brackets mean that they are fixed.
The formal errors given by GALFIT are all tiny:
$<0.05$ for magnitude and \sersic\ index, $<0.\arcsec1$ for $r$.}
\label{tbl-galfit}
\end{deluxetable}

\begin{deluxetable}{lllccccccc}
\tabletypesize{\scriptsize}
\tablecaption{\scshape Photometric Data of \src}
\tablewidth{0pt}
%\tablehead{\colhead{Instrument} & \colhead{Wavelength} &\colhead{Flux Density} & \colhead{Unit} & Reference }
\tablehead{\colhead{Instrument} & \colhead{Wavelength or Energy} &\colhead{log ($\nu L_{\nu}$/\erg)} & Reference }

%\hspace*{15.mm}    &  10$^{-12}$ erg cm$^{-2}$ s$^{-1}$  &(keV)  &(eV)    & 10$^{-5}$ photons cm$^{-2}$ s$^{-1}$ &  $\rm ergs\ s^{-1}$   & $\rm ergs\ s^{-1}$  & $\rm ergs\ s^{-1}$}
%\hspace*{15.mm} (1)&(2) & (3) & (4) & (5) & (6) & (7) & (8)   }
\startdata
%\hline
GALEX$^\dag$  &  1539\AA~(FUV)  & 41.86 (41.33)   & (1) \\
GALEX$^\dag$  &  2316\AA~(NUV)  & 42.29 (41.06)   & (1) \\
XMM$^\dag$    &  2120\AA~(UVW2) & 41.83 (41.98)   & (1) \\ 
XMM$^\dag$    &  2910\AA~(UVW1) & 42.39 (41.74)   & (1) \\
SDSS  &  3588\AA~($u$)       & 42.88  & (2) \\
SDSS  &  4862\AA~($g$)       & 43.31  & (2) \\
SDSS  &  6289\AA~($r$)       & 43.39 & (2) \\
SDSS  &  7712\AA~($i$)       & 43.40 & (2) \\
SDSS  &  9230\AA~($z$)       & 43.38  & (2) \\
HST$^\dag$    &  4500\AA~(B)        &  43.19 (41.90)& (1) \\
HST$^\dag$    &  8140\AA~(I)        & 43.35 (41.83) & (1) \\
2MASS &  1.22\ums  (J)        & 43.37  & (2) \\
2MASS &  1.61\ums  (H)        & 43.31 & (2) \\
2MASS &  2.12\ums  (K)        & 43.14 & (2) \\
WISE  &  3.4 \ums (W1)             & 42.51 & (1) \\
WISE  &  4.6 \ums (W2)            & 42.12 & (1) \\
WISE  &  12 \ums  (W3)           & 41.46 & (1) \\
WISE  &  22 \ums  (W4)           & $<$41.70 & (1) \\
Spitzer & 24\ums (MIPS)     & 41.22   &  (3) \\
VLA &  20 cm    & 37.19 & (4) \\
XMM &  0.5--2 keV (PN quiescent sate)  & 41.83  & (5) \\
XMM &  0.5--2 keV (PN flare sate)  & 40.45  & (5) \\
XMM &  2--10 keV (PN)  & $<$40.31  & (5) \\
\enddata
\tablecomments{(1) This work; (2) From NED database; (3) Mahajan et al. (2010); (4) Miller et al. (2009); 
(5) Sun et al. (2013). $^\dag$Host galaxy flux from the GALFIT decomposition, while the AGN flux is given in parenthesis.  
%; (4) Barger et al. 2014; (5) Chapin et al. 2009.
}
\end{deluxetable}

 \begin{table} \caption{Best Fit {\sc optxagnf} Model \label{tab:bestfit_neut}}
\begin{center}
\begin{tabular}{llccc}
\hline
\hline
 Model Component  & Parameter & Quiescent  &  Flare \\
 \hline
&&   \\
Extinction         & E(B-V)             &  0.0128$^\dag$  & 0.0128$^\dag$ \\
Absorption         & $N_{\rm H}$ ($10^{20}$ cm$^{-2}$) &   $1.08
(<1.7)$ & $1.62 (<3.3)$  
 \\
%& Normalisation & $3.82_{-0.48}^{+0.63}\times 10^{-3}$ \\
optxagnf  & $M_{\rm BH}$(\msun) & 2.8$^{+1.0}_{-0.7}\times10^6$  & 1.7$^{+3.5}_{-0.8}\times10^6$\\
          & log$L/L_{\rm Edd}$  & -0.84$^{+0.12}_{-0.13}$ & -0.82$^{+0.29}_{-0.48}$ \\
          & $r_{\rm cor}$ ($R_{\rm g}$) & 10$^\dag$  &  21.2$(>11.8)$ \\ 
          & log$R_{out}$ ($R_{\rm g}$) & 5.0$^\dag$  &  5.0$^\dag$ \\
          &  $kT_{e}$ (keV)$^\star$    &   0.1$^{+1.0}_{-0.04}$ & 0.13$^{+0.11}_{-0.03}$ \\
          &  $\tau$$^\star$:           &   18.0$^{+38}_{-18.0}$ & 24.2$^{+49}_{-6.4}$ \\
          & $f_{\rm pl}$      &   0.07$^{+0.06}_{-0.07}$  &  $0.18^{+0.30}_{-0.18}$\\
 \hline
Statistics          &  $\chi^2/dof$    & 26.4/40 & 37.9/24\\
%    &$\chi^2/dof$&454/406 & \\

%L$_{(2-10)\mathrm {keV}}^{\rm b}$  & Unabsorbed~(\lum) & $5.8 \times 10^{42}$ & $8.1 \times 10^{42}$ & $1.04 \times 10^{43}$\\
%Statistics$^\ast$  & $\chi^2/dof$&391.7/313 (390.6/313) & 417.8/314 (410.5/314) & 404.1/308 (397.6/308)\\
 \hline
\end{tabular}
\end{center}
%\begin{center}
{\bf Note--}
$^\star$: The 90\% confidence errors are calculated with BH mass fixed.  
$^\dag$: The parameters are fixed during the fittings. 
For the flare state spectral fittings, the errors on BH and log$L/L_{\rm Edd}$  are
derived by fixing the corona breaking radius. \\
%$^\ddag$: A meaningful constraint on the upper limit of iron abundance cannot be obtained. \\  
%\end{center}
\end{table}

\begin{table}
\caption{ Low Mass AGNs with Extremely Soft X-ray Emission}
\centering
\begin{tabular}{llllllllllc}
%\small
\hline\hline
Name & Morph. & $z$ & log($M_{\rm BH}$) & $\Gamma$$^a$ & log(L$_{0.5-2keV}$$^a$) & log(L$_{O[III]}$) &
log(L$_{1.4\rm GHz}$) & $\alpha_{\rm ox}$ & $\Delta\alpha_{\rm ox}^b$ & Ref. \\
& & & \msun  &  & \erg &  \erg &  \erg  &  &   & \\ 

%   &  &        &                  & 10$^{20}$ & 10$^{-14}$  & & &10$^{-3}$ & & \\
%   &  &        &                  & cm$^{-2}$ & erg cm$^{-2}$ s$^{-1}$ & & &cts/s & &\\
\hline
RX J1301-2746 & disk & 0.024 & 5.9 & 7.1(4.4) & 40.45 (41.8) & 39.64 & 37.6 &  -2.04 (-1.50) & -0.99 (-0.46) &1 \\
2XMM J1231+1106 & disk & 0.119 & 5 & 4.8      & 42.13 (42.5) & 40.22 & $\dots$ & -1.89 (-1.68) & -0.77 (-0.57) &2,3 \\
GSN 069  & $\dots$  & 0.018 & 6.08 & 6.7 & 42.08 & 40.32 & $\dots$ &$<$-2 & -0.89 &4\\
%23 18 54.8 & -42 14 20.6        &.005& 40.79 &  & 1.7 & 0.55 & 0.86 & 1.40 & 2.57 & 4.96  \\  
%13:02:00.1&+27:46:57.6 & 0.0237 & 3.2 & \dots &  2   & 3.$\pm$0.2?  & 33$\pm3?$  & 2$^{+?}_{-?}$ & ?  \\
%13:02:00.1&+27:46:57 & 0.024 & 2.9 & 1.5 &  2   & $3.9^{+0.7}_{-0.4}$ & 26$\pm3$  & 1.0$^{+1.1}_{-1.4}$ & 21.3  \\
%& & &2.9 & 0.57  & 2.5 & $3.2\pm0.2$ & 30$\pm2$  & $\dots$     & 2.9  \\
%& & &2.9 & 0.34  &  3  & $3.6\pm0.2$ & 28$^{+3}_{-2}$  & $\dots$     & 2.3  \\
%
%
%& & & &   &  2.5   &   &   &  &  \\ 
%& & & &   &  3.0   & 3.6$\pm$0.2  & 28$^{+3}_{-2}$  & \dots & 2.2 \\  
\hline\\
\end{tabular}
\vspace{-0.1cm}
%\\$^a$ All quantities in cgs units\\
%\tablenote
%    \begin{tablenotes}
\scriptsize
%{\\$^a$ flux in 10$^{-14}$ erg cm$^{-2}$ s$^{-1}$;
%$^b$ $\Gamma_h$ is the assumed hard X-ray photon indices in the 2-7 keV, 
%while $\Gamma_1$ and $\Gamma_2$ are indices in the soft and hard band for the simulated spectra;  
{\\$^a$ Parentheses show the values corresponding to flare state.
{$^b$ The difference between $\alpha_{\rm ox}$ and the value expected from the 
\citet[][]{steffen06} $\alpha_{\rm ox}-L_{\rm 2500\AA}$ relation.}
(1) S13; (2) Ho et al. 2012; (3) Terashima et al. (2012); (4) Miniutti et al. (2013)
%represents the significance of the second power-law ($\Gamma_2$) in the simulated spectra.
}
%    \end{tablenotes}
\vspace{-0.5cm}
\end{table}

\clearpage
\newpage

\appendix
%\begin{deluxetable}{c|cccccccc}
\section{Test on the \hst~AGN-to-galaxy decomposition}

\begin{figure}
\centering
{
\includegraphics[width=0.8\textwidth]{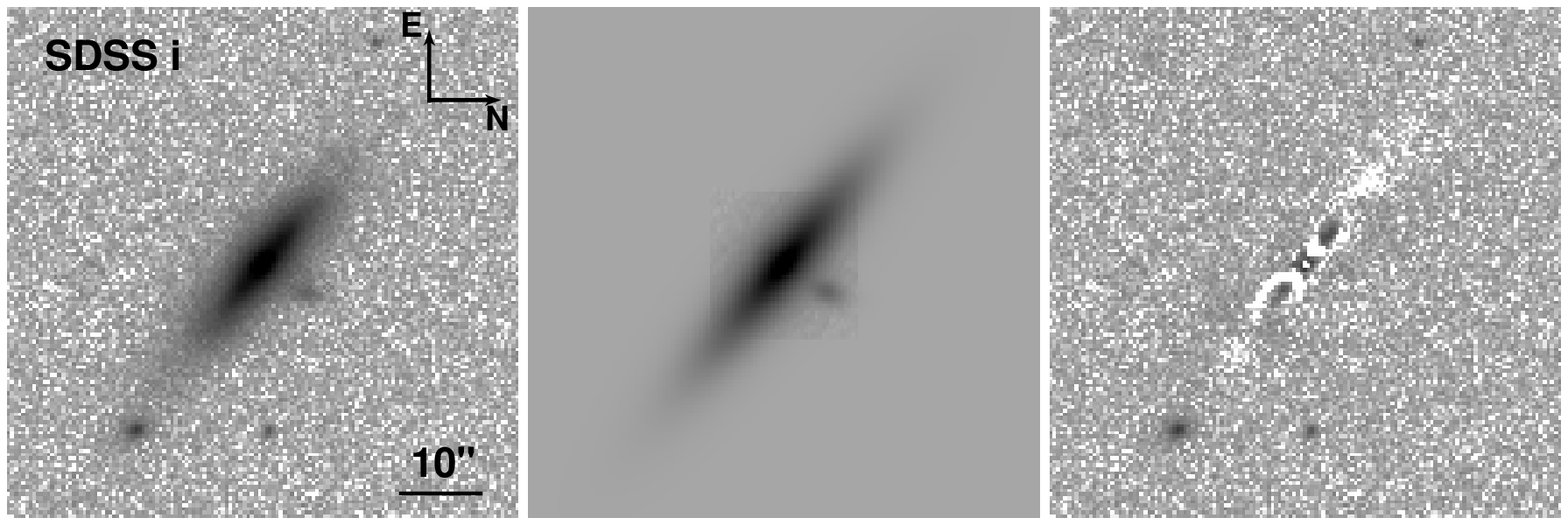}%
\hfil
\includegraphics[width=0.5\textwidth]{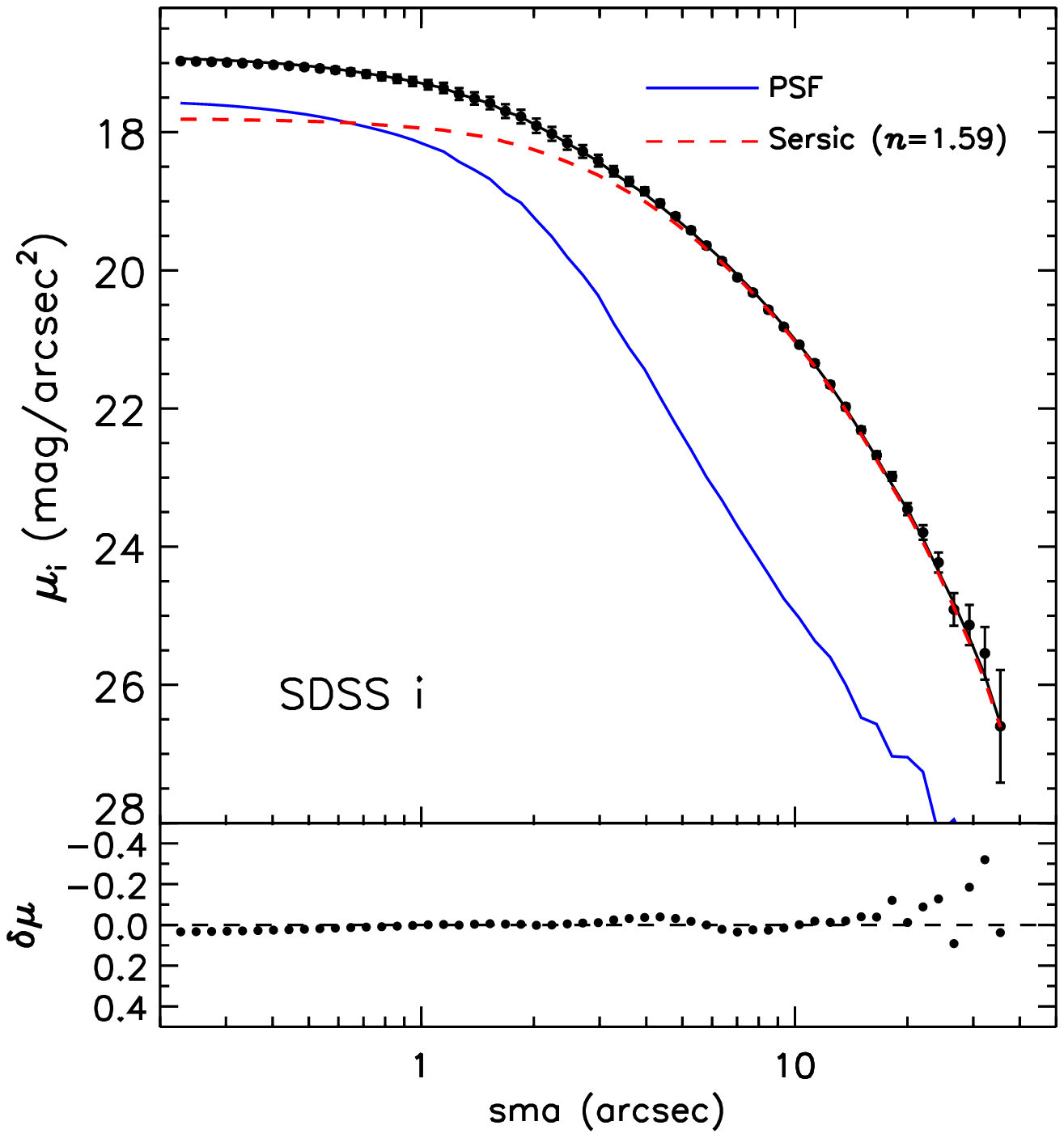}%
%\hfil
}
\caption{ One-dimensional surface brightness distribution the SDSS I-band image, 
along with that for the best-fit PSF (blue solid line) and the \sersic~component (red dashed).  
 }
\end{figure}

{Due to the complex galaxy morphology observed with \hst, our GALFIT decompositions
may introduce potential systematic errors. 
%However, quantifying the effect of systematic uncertainties
%is not a trivial task. 
We have looked at images and identified the possible bar structures, which can be
seen in Figure 1. In fact, our fittings with PSF+sersic models are quiet good, except for the large
residuals at large radius ($r>5$ arcsec). The disk component is always fitted by an exponential profile,
as in nearby inactive galaxies and galaxies with IMBHs, although disk profiles do vary at large radii 
\citep[][]{jiang11b}. 
Therefore, we investigated the effect of possible systematic biases due to the variations of outer disk profile. 
For simplicity, we choose to work on the \hst~F814W image. 
We first allowed the \sersic~index for the outer disk component
to vary in our fits, but found similar results. 
This is perhaps because \hst~is not able to detect extended, lower surface brightness emission component of the 
galaxy when the exposure is not deep \citep[e.g.,][]{vandokkum15}. 
We then turned to the SDSS I-band image which has a larger field of view 
with very good measurements of the sky background \citep[][]{gunn98}. 
This is important in studying the extended galaxy morphology,
particularly the low-surface-brightness structure like galactic disk 
\citep[e.g.,][]{pohlen06, erwin08, jiang13}.
%(e.g., Pohlen \& Trujillo 2006,A\&A,454,759; Erwin et al. 2008,AJ,135,20; Jiang et al. 2013;ApJ,770,3).
We found the disk can be modeled with a \sersic~ component but with a
slightly higher index of $n=1.59$. There are no systematic residuls
left as can be see from the one-dimensional surface brightness
distribution (Figure A1). 
We then fixed the \sersic~index at $n=1.59$ and refitted the model to
the HST F814W image.  We found most parameters remain unchanged in our fits, except for the
magnitudes.  Though the residuals at large radii disappear (Figure A2, right panel),
the fit becomes worse in the inner region where the PSF and central
bulge dominate. A comparison of results by fixing $n=1$ (exponential
profile) and  $n=1.59$ can be found in Figure A2 and Table A1. 
Since the difference between magnitudes of the AGN and galaxy
component is less than 0.3 mag, we conclude that our fits are not
introducing major systematic errors.

%we found that the residuals at large radius can be modelled with a slightly higher \sersic~index (n=1.59).  
 }

%\begin{deluxetable}{ccccccccc}
%\tiny
%\tabletypesize{\small}
%\tablewidth{0pt}
%\tablecaption{GALFIT Decomposition of \hst\ Images\label{tbl-galfit}}
%\tablenum{2 . GALFIT Decomposition}
%\tablehead{ \colhead{Filter} & \colhead{Component} & \colhead{$m$} & \colhead{$M$}
%& \colhead{$n$} & \colhead{$r$ (\arcsec/pc)} & \colhead{$b/a$} & \colhead{$c$} \\
%(1) & (2)    & (3)  & (4) & (5) & (6) & (7) & (8) }
%\startdata
\begin{table}
\caption{GALFIT Decomposition of \hst\ Images\label{tbl-galfit}}
\centering
\begin{tabular}{llllccllllc}
%\small
\hline\hline
%Filter & Component & $m$  &$f_{\lambda}$ &$n$ & $r$(\arcsec/kpc) & b/a   & PA & $\chi^2/dof$\\
Filter & Component & $m$ & $M$ & $n$ & $r$ (\arcsec/pc) & $b/a$ & $c$ \\
(1)    &  (2)            &  (3)       &  (4)  &  (5)       &  (6) & (7)  & (8) \\
\hline
F814W (old)  & PSF      & 18.12 & $-16.81$ & \nodata   & \nodata    & \nodata & \nodata   \\
       & \sersic\ & 16.39 & $-18.67$  & 1.45  & 0.41/193      & 0.54    & 2.63 \\
       & Exp Disk & 14.50 & $-20.56$  & [1.0] & 2.86/1332     & 0.24  & \nodata     \\
F814W (new) & PSF      & 17.98 & $-16.95$ & \nodata   & \nodata    & \nodata & \nodata   \\
       & \sersic\ & 16.73 & $-18.33$  & [1.45]  & [0.41/193]      & 0.61    & 5.00  \\
       & \sersic\ & 14.33 & $-20.73$  & [1.59] & 4.97/2338     & 0.25  & \nodata     \\
%F450W  & PSF      & 19.06 & $-16.04$  & \nodata &  \nodata    & \nodata   & \nodata \\
%       & \sersic\ & 17.56 & $-17.35$  & [1.43]  & 0.36/174    & 0.52 &  1.42        \\
%       & Exp Disk & 16.04 & $-18.90$  & [1.0]   & 2.71/1300   & 0.24 & \nodata   \\
%\enddata
%\tablecomments{ Col. (1): $HST$ filter.
\hline\\
\end{tabular}
\vspace{-0.1cm}
{
\\
{Note--}  Col. (2): Components used in the fitting schemes.
Col. (3): The integrated magnitudes on the Vega system, not corrected for
Galactic extinction.
Col. (4): The absolute Johnson $I$ and $B$ magnitude after Galactic extinction
correction. We assume a power-law continuum ($f_\nu \propto \nu^{-0.5}$
for the central AGN, an Sb galaxy for the \sersic\ component,
and an $S0$ galaxy for the disk, using templates from Kinney et al (1996).
Col. (5): The \sersic\ index.
Col. (6): The effective radius of the \sersic\ component or scale length of
exponential disk, in units of arcsec and pc, respectively.
Col. (7): Axis ratio.
Col. (8): Diskiness (negative)/boxiness (positive) parameter,
defined in Eqn.~(3) of Peng et al. (2002).
The brackets mean that they are fixed.
The formal errors given by GALFIT are all tiny:
$<0.05$ for magnitude and \sersic\ index, $<0.\arcsec1$ for $r$.
}
\vspace{1.cm}
\end{table}

\begin{figure}
\centering
{
\includegraphics[scale=0.7, angle=0]{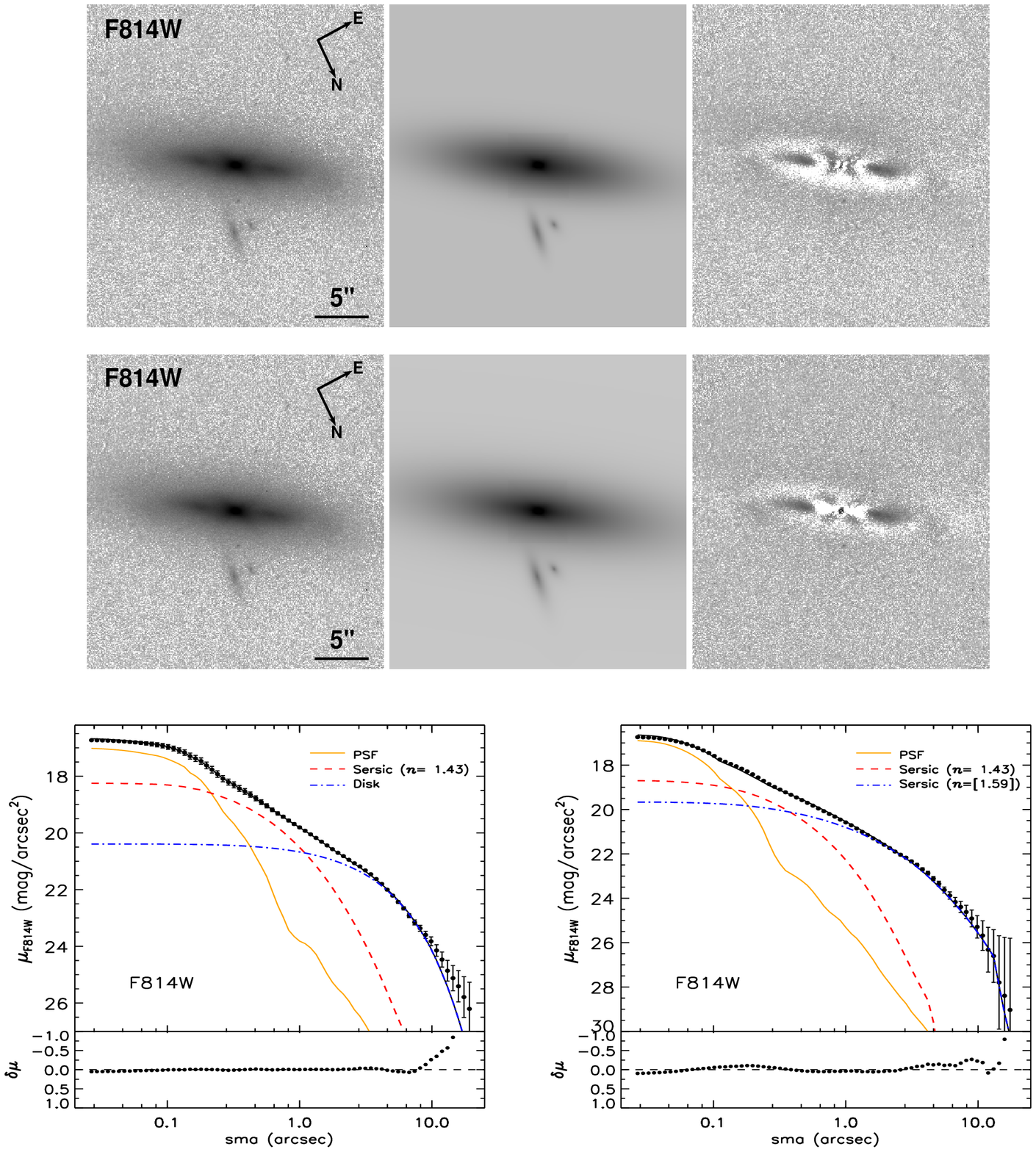}
}
\caption{Galfit decomposition of the \hst~F814W image of J1302 by 
fixing the index of disk component at $n=1$ (exponential
profile, upper) and $n=1.59$ (lower).  
 A comparison of the one-dimensional surface brightness distribution 
 of the two is shown in the bottom panel. 
}
\end{figure}

%residuals are still
%present, and the fit statistics is not improved. 

\section{Test on the AGN-to-galaxy decomposition in the UV}

\begin{figure}
\centering{
\includegraphics[scale=0.8, angle=0]{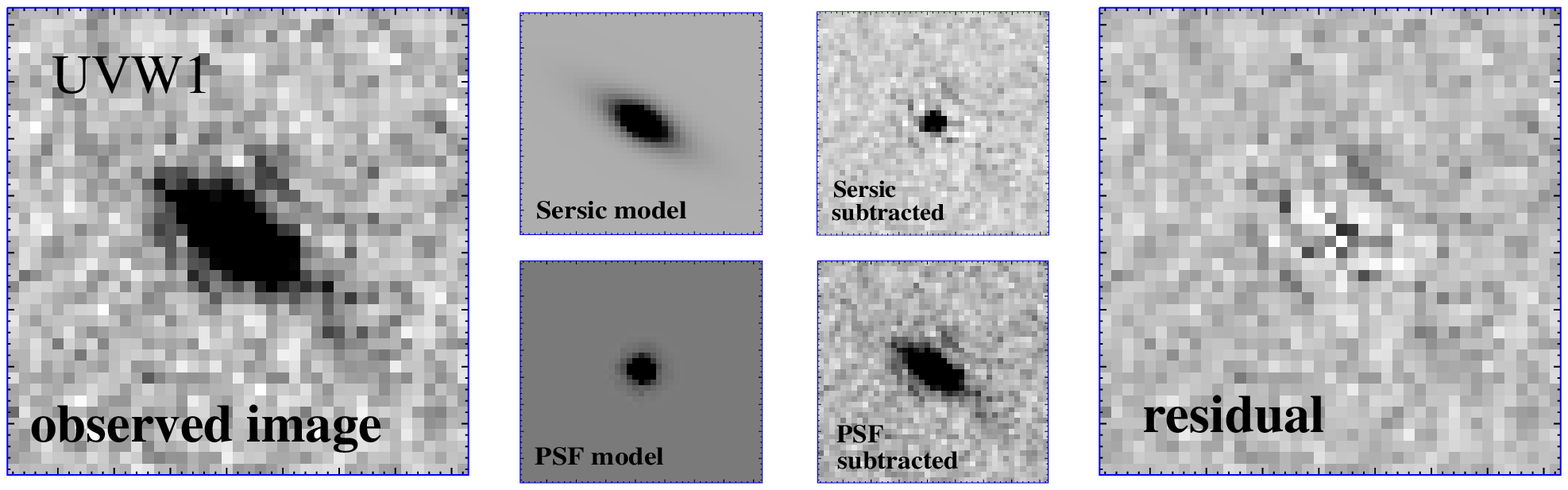}
\caption{
%UV images of J1302. The image in the upper row takes from \xmm/OM
%at wavelength $\lambda_{\rm eff}=2120$\AA (UVW2) and $\lambda_{\rm eff}=2910$\AA (UVW1),
%while lower row shows, from left to right, the image in GALEX FUV (1539\AA) and NUV
%(2316\AA), respectively.
%It is clear that the galaxy in the shorter UV wavelength is dominated by a point source
%emission, possibly associated with an AGN.
%Each image has a size of $1\times1$ arcmin$^2$.
\xmm-OM UVW1 Host-to-AGN decomposition of \src. 
The first column shows the observed image of the galaxy. 
The second column shows the best-fit galaxy (top, parameterized by a single 
\sersic~profile) and AGN (bottom, parameterized by a single PSF) 
components as extracted by GALFIT. 
The third column shows the corresponding residual after subtraction of the \sersic~ 
(PSF).  
The residual image after both components being subtracted is shown on the right.  
}\label{fig:galfit_uvw1}}
\end{figure}

\begin{figure}
\centering{
\includegraphics[scale=0.6, angle=0]{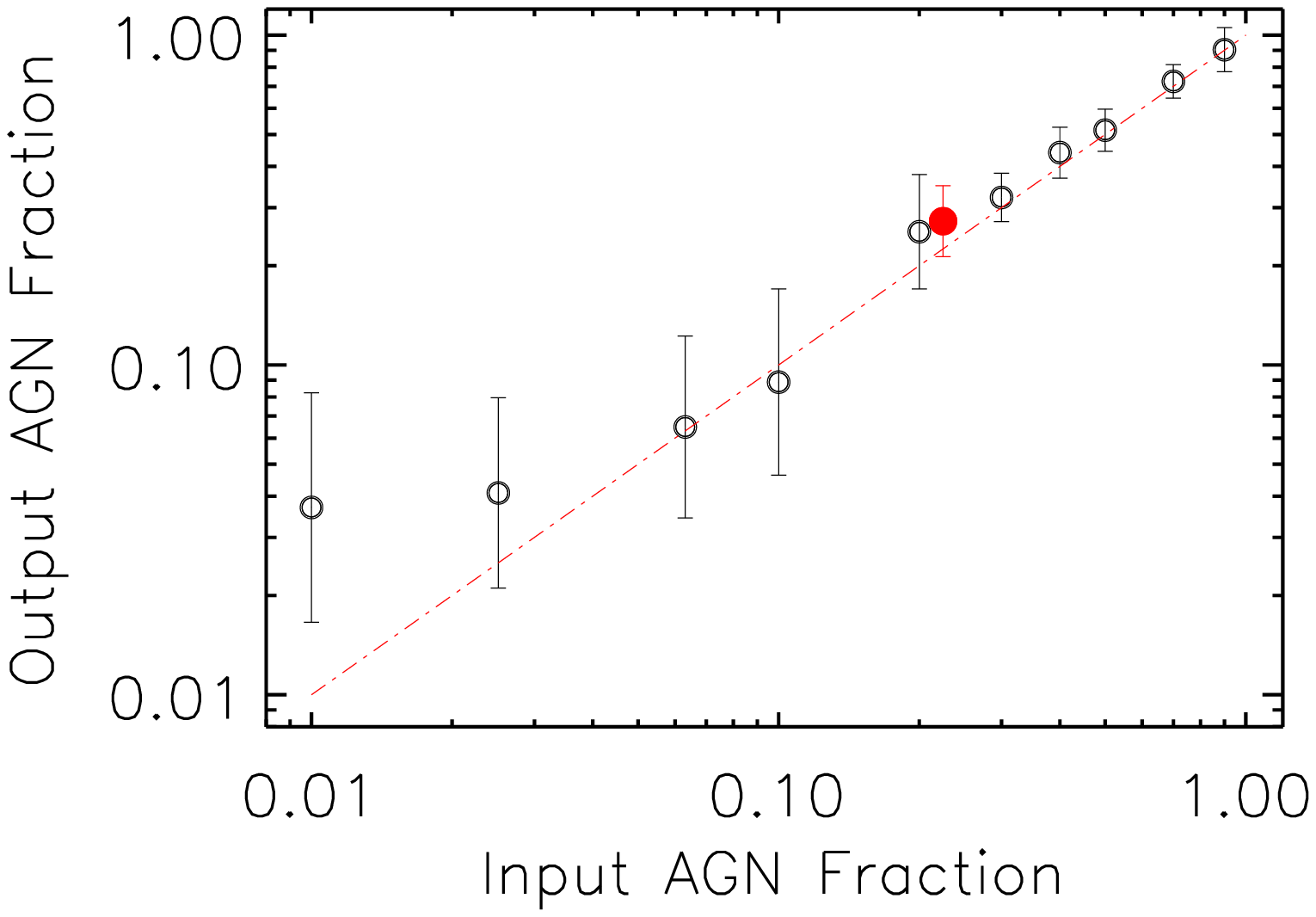}
\caption{
Comparison between the simulated AGN-to-galaxy ratio and that measured by GALFIT. 
The red filled circle represents the best-fit AGN fraction of 22.5\% for \src in the UVW1.  
The dot-dashed line is the one-to-one relation. 
%The vertical dotted lines correspond to the FWHM of H$\beta$ line (Wang et al. 2007).}
}\label{fig:simulate_fuv}}
\end{figure}

\begin{figure}
\centering{
\includegraphics[scale=0.8, angle=0]{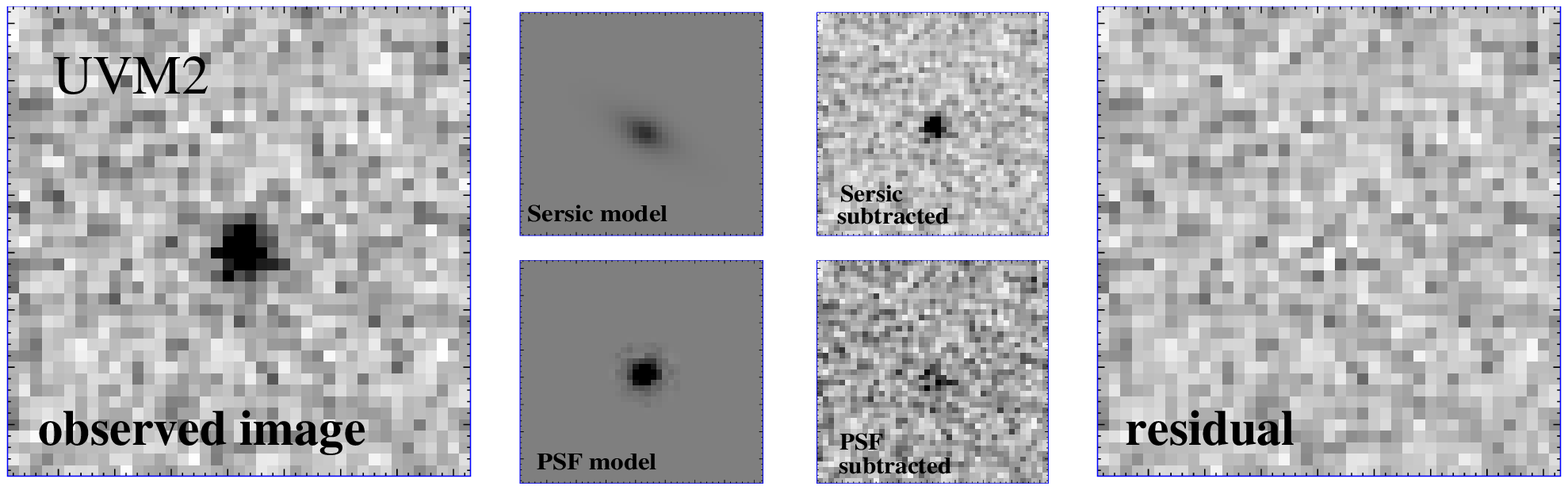}
\caption{
The same as Figure A1 but for the OM UVW2 filter. 
%The vertical dotted lines correspond to the FWHM of H$\beta$ line (Wang et al. 2007).}
}\label{fig:galfit_uvm2}}
\end{figure}

%\section{Test on the AGN-to-galaxy decomposition in the UV}
{   
Having established the central point source and pseudobulge detections in the \hst~optical, 
we try to perform similar image decomposition of the AGN and host galaxy of \src 
in the UV using GALFIT. Although the UV images have a low spatial resolution ($\sim2$\arcsec) 
compared with the \hst~optical, they are still very useful for us to understand its host 
galaxy and put further constraint on the SED of the AGN (Section 3.1). 
% by GALFIT decomposition, 
%we sought to check whether there is a point-like emission in the UV images taken with \xmms/OM. 
%Erwin et al. 2008; Li et al. 2011).
We chose to first work with the \xmm OM UVW1 image ($\lambda_c=2910$\AA), which is closest to the HST F450W filter and has a better resolution than the GALEX. 
The UVW1 PSF is constructed from the field of 3C273 by stacking 5 bright, 
isolated point sources from the catalog of Page et al. (2012). 
As for the \hst, the host galaxy is represented by a \sersic~profile 
and we performed the fittings with PSF+\sersic~allowing all parameters to vary. 
The image decomposition is shown in Figure~\ref{fig:galfit_uvw1}. 
The fit is good with a reduced $\chi^2/dof=1.1$, and the best-fit parameters for the galaxy, i.e., \sersic~index, axis ratio, and position angle, are all close to that obtained from the \hst. 
Note that fitting with a {\it single} \sersic~profile leaves acceptably residuals, 
but the fit statistics is worse (the difference of $\chi^2$ is 13, Table A1).
In addition, the \sersic~index ($n=4.28$) is much larger than that obtained from the 
\hst~optical ($n=1.45$), suggesting that a more concentrated light distribution is 
required which may be due to the presence of unresolved AGN emission.  
%and statistics is worse (the difference of $\chi^2$ is 13, Table ???).  

%Fixing the \sersic index to 
%and the fitting 
%results are given in Table ??.  
%For comparison, we also performed the GALFIT fittings by fixing the \sersic 

Since the \xmms/OM spatial resolution is much worse than the \hst, we 
sought to test the robustness of our AGN+galaxy image decomposition. 
To do so, we used GALFIT to simulate a point source along with a \sersic~profile host 
galaxy, with total flux adding up to the measured flux in the OM/UVW1. 
% AGN fraction changing 
%from 0.1 to 0.9.  
%We then performed the same analysis that we used on the real
%quasar image, w
We create a large set of simulated galaxies with profiles assumed to be the 
best fit \sersic~ model obtained above, varying the total integrated flux of the central 
point source from $m$=24.7 to 19.7 mag. The magnitude range represents point sources 
with fluxes from $\simeq$0.9 to 1/100 of the host galaxy flux. 
%Fainter host galaxies than this are undetectable due to shot noise from the point source. 
We then place these models randomly on empty regions of the real UVW1 image, trying 
to measure their properties in the presence of photometric noise. 
Note that when the AGN emission exceeds the 70\% of the host galaxy, 
the fitting yields an unphysical high value of \sersic~index ($n>$10). 
In this case, we set constraints on the \sersic~index $n$ to values within 0.7 to 6 
to avoid catastrophic fitting results. 
Figure~\ref{fig:simulate_fuv} summarizes the results of these simulations, plotting 
the input AGN fraction against the measured fraction from the simulated images.  
We find that the decomposition method is able to recover the AGN flux with 
a $\sim$80\% accuracy at $m_{\rm AGN}\simgt22.2$, corresponding to the 
ten percent of the galaxy flux. Below $m_{\rm AGN}\simgt22.2$, the fittings tend to overestimate 
the AGN flux. For instance, the measured AGN fraction is a factor of four higher than the 
input value at $m_{\rm AGN}\sim24.7$.
%A similar AGN and galaxy image decomposition 
We also performed the AGN and galaxy decomposition for the \xmms/OM UVW2 image ($\lambda_c=2120$\AA), 
which is shown in Figure~\ref{fig:galfit_uvm2}. 
Since the galaxy is marginally resolved in the UVW2, 
we fit its profile with the same \sersic~ model as that of UVW1 image, but 
with all parameters held fixed except its magnitude. 
The results of the decomposition are shown in Table A1. 
Note that the total flux (AGN+\sersic) from our GALFIT decomposition in the both filters 
is in good agreement with the aperture photometry from the \xmms/OM UV catalog (Page et al. 2012).   
}

%It should be noted that these simulations are only be able to 
%the ability
%of the codes to recover what was put on the simulated image, i.e.,
%idealized profiles with realistic photometric noise, but it does not allow us to say how reliable is
%the decomposition in the case of perturbed, irregular or clumpy

\begin{table}
\caption{Galfit Decomposition of OM UV images}
\centering
\begin{tabular}{llllccllllc}
%\small
\hline\hline
Filter & Component & Magnitude  &$f_{\lambda}$ &$n$ & $r$(\arcsec/kpc) & b/a   & PA & $\chi^2/dof$\\
(1)    &  (2)            &  (3)       &  (4)  &  (5)       &  (6) & (7)  & (8) & (9)\\
%& & & $10^{-14}$ \ergs & $10^{-14}$ \ergs &   & eV  &  & \\ 
%   &  &        &                  & 10$^{20}$ & 10$^{-14}$  & & &10$^{-3}$ & & \\
%   &  &        &                  & cm$^{-2}$ & erg cm$^{-2}$ s$^{-1}$ & & &cts/s & &\\
\hline
UVW1 2910\AA& PSF & 21.29$\pm0.19$ & 1.45$\pm0.44$ & \dots & \dots & \dots & \dots  & \dots \\
                         & \sersic & 19.67$\pm$0.06 & 6.45$\pm0.68$ & 2.06$\pm$0.37 & 3.57$\pm$0.16/(1.68$\pm0.08$) & 0.3$\pm$0.01 & 57.3$\pm$0.8  & 1861/1670 \\
\hline
                         & Single \sersic & 19.37$\pm$0.03 & 8.50$\pm0.23$ & 4.28$\pm$0.4 & 2.75$\pm0.14$/(1.29$\pm0.07$) & 0.3$\pm$0.01 & 57.3$\pm$0.8 & 1874.1/1673 \\
\hline
UVW2 2120\AA& PSF & 23.04$\pm0.07$ & 3.47$\pm0.58$  & \dots & \dots & \dots & \dots & \dots \\
                        & \sersic$^\dag$ & 23.42$\pm0.2$ & 2.45$\pm0.79$ & [2.06] & [3.75] & [0.3] & [57.3] & 1792/1674 \\ 
%23 18 54.8 & -42 14 20.6        &.005& 40.79 &  & 1.7 & 0.55 & 0.86 & 1.40 & 2.57 & 4.96  \\  
%13:02:00.1&+27:46:57.6 & 0.0237 & 3.2 & \dots &  2   & 3.$\pm$0.2?  & 33$\pm3?$  & 2$^{+?}_{-?}$ & ?  \\
%13:02:00.1&+27:46:57 & 0.024 & 2.9 & 1.5 &  2   & $3.9^{+0.7}_{-0.4}$ & 26$\pm3$  & 1.0$^{+1.1}_{-1.4}$ & 21.3  \\
%& & &2.9 & 0.57  & 2.5 & $3.2\pm0.2$ & 30$\pm2$  & $\dots$     & 2.9  \\
%& & &2.9 & 0.34  &  3  & $3.6\pm0.2$ & 28$^{+3}_{-2}$  & $\dots$     & 2.3  \\
%
%
%& & & &   &  2.5   &   &   &  &  \\ 
%& & & &   &  3.0   & 3.6$\pm$0.2  & 28$^{+3}_{-2}$  & \dots & 2.2 \\  
\hline\\
\end{tabular}
\vspace{-0.1cm}
%\\$^a$ All quantities in cgs units\\
%\tablenote
%    \begin{tablenotes}
\scriptsize
%{\\$^a$ flux in 10$^{-14}$ erg cm$^{-2}$ s$^{-1}$;
%$^b$ $\Gamma_h$ is the assumed hard X-ray photon indices in the 2-7 keV, 
%while $\Gamma_1$ and $\Gamma_2$ are indices in the soft and hard band for the simulated spectra;  
{
\\
{Note--} Col. (1): \xmm OM filter.
Col. (2): Components used in the fitting schemes.
Col. (3): The magnitudes measured from the UV counts assumed that 
the photometric zero point is $m_{\rm 0}=20$, not corrected for
Galactic extinction. 
Col. (4): Flux density in units of $10^{-16}$\ergs$\AA^{-1}$. 
Col. (5): The \sersic\ index.
Col. (6): The effective radius of the \sersic\ component, in units of arcsec and kpc, respectively.
Col. (7): Axis ratio.
Col. (8): Position angle (degree). 
Col. (9): The fit statistics.   
$^\dag$The brackets mean that they are fixed.
%represents the significance of the second power-law ($\Gamma_2$) in the simulated spectra.
}
%    \end{tablenotes}
\vspace{0.5cm}
\end{table}

%\begin{figure}
%\centering{
%\includegraphics[scale=0.6, angle=0]{images/j1302_UVimage2.ps}
%\includegraphics[scale=0.5, angle=0]{/Users/mac/paper/j1302/sed/submit/160723/images/xmmom/simulate/fuv/figure/fig_psf_UVM2.eps}
%\caption{
%The same as Figure?? but for the OM UVM2 filter.
%The vertical dotted lines correspond to the FWHM of H$\beta$ line (Wang et al. 2007).}
%}\label{fig:psf_uvm2}}
%\end{figure}


\begin{thebibliography}{}
\bibitem[Ai et al.(2011)]{ai11} Ai, Y.~L., Yuan, W., Zhou, H.~Y., Wang, T.~G., \& Zhang, S.~H.\ 2011, \apj, 727, 31 
\bibitem[Antonucci(1993)]{1993ARA&A..31..473A} Antonucci, R.\ 1993, \araa, 31, 473 
%\bibitem[Bai et al.(2006)]{bai06} Bai, L., Rieke, G.~H., Rieke, M.~J., et al.\ 2006, \apj, 639, 827 

\bibitem[Begelman et al.(2006)]{begelman06} Begelman, M.~C., Volonteri, M., \& Rees, M.~J.\ 2006, \mnras, 370, 289 

\bibitem[Bekki et al.(2006)]{2006ApJ...642L.133B} Bekki, K., Couch, W.~J., \& Shioya, Y.\ 2006, \apjl, 642, L133 
\bibitem[Bentz et al.(2009)]{bentz09} Bentz, M.~C., Peterson, B.~M., Pogge, R.~W., \& Vestergaard, M.\ 2009, \apjl, 694, L166 

\bibitem[Biggs \& Ivison(2006)]{biggs06} Biggs, A.~D., \& Ivison, R.~J.\ 2006, \mnras, 371, 963 

\bibitem[Caldwell \& Rose(1997)]{1997AJ....113..492C} Caldwell, N., \& Rose, J.~A.\ 1997, \aj, 113, 492 
\bibitem[Caldwell et al.(1999)]{caldwell99} Caldwell, N., Rose, J.~A., \& Dendy, K.\ 1999, \aj, 117, 140 
\bibitem[Chary \& Elbaz(2001)]{ce01} Chary, R., \& Elbaz, D.\ 2001, \apj, 556, 562 

\bibitem[Chartas et al.(2009)]{chartas09} Chartas, G., Kochanek, C.~S., Dai, X., Poindexter, S., \& Garmire, G.\ 2009, \apj, 693, 174 

\bibitem[Corbel et al.(2013)]{corbel13} Corbel, S., Coriat, M., Brocksopp, C., et al.\ 2013, \mnras, 428, 2500 

\bibitem[Crummy et al.(2006)]{crummy06} Crummy, J., Fabian, A.~C., Gallo, L., \& Ross, R.~R.\ 2006, \mnras, 365, 1067 

\bibitem[Desroches et al.(2009)]{desroches09} Desroches, L.-B., Greene, J.~E., \& Ho, L.~C.\ 2009, \apj, 698, 1515 
\bibitem[Dewangan et al.(2008)]{dewangan08} Dewangan, G.~C., Mathur, S., Griffiths, R.~E., \& Rao, A.~R.\ 2008, \apj, 689, 762-774 
\bibitem[Done et al.(2007)]{done07} Done, C., Gierli{\'n}ski, M., \& Kubota, A.\ 2007, \aapr, 15, 1 
\bibitem[Done et al.(2012)]{done12} Done, C., Davis, S.~W., Jin, C., Blaes, O., \& Ward, M.\ 2012, \mnras, 420, 1848 
\bibitem[Dong et al.(2012)]{dong12} Dong, X.-B., Ho, L.~C., Yuan, W., et al.\ 2012, \apj, 755, 167
\bibitem[Dong et al.(2012)]{dong12} Dong, R., Greene, J.~E., \& Ho, L.~C.\ 2012, \apj, 761, 73 
\bibitem[Elbaz et al.(2010)]{elbaz10} Elbaz, D., Hwang, H.~S., Magnelli, B., et al.\ 2010, \aap, 518, L29 
\bibitem[Elitzur \& Ho(2009)]{elitzur09} Elitzur, M., \& Ho, L.~C.\ 2009, \apjl, 701, L91 
\bibitem[Elvis et al.(1994)]{elvis94} Elvis, M., Wilkes, B.~J., McDowell, J.~C., et al.\ 1994, \apjs, 95, 1  

\bibitem[Erwin et al.(2008)]{erwin08} Erwin, P., Pohlen, M., \& Beckman, J.~E.\ 2008, \aj, 135, 20 

\bibitem[Fabian et al.(2009)]{fabian09} Fabian, A.~C., Zoghbi, A., Ross, R.~R., et al.\ 2009, \nat, 459, 540 
\bibitem[Falcke et al.(2004)]{falcke04} Falcke, H., K{\"o}rding, E., \& Markoff, S.\ 2004, \aap, 414, 895 \
\bibitem[Ferrarese \& Merritt(2000)]{Ferrarese00} Ferrarese, L., \& Merritt, D.\ 2000, \apjl, 539, L9 
\bibitem[Filippenko \& Ho(2003)]{filippenko03} Filippenko, A.~V., \& Ho, L.~C.\ 2003, \apjl, 588, L13 
\bibitem[Fukugita et al.(1995)]{fukugita95} Fukugita, M., Shimasaku, K., \& Ichikawa, T.\ 1995, \pasp, 107, 945 
\bibitem[Gallo et al.(2012)]{gallo12} Gallo, E., Miller, B.~P., \& Fender, R.\ 2012, \mnras, 423, 590 
\bibitem[Gebhardt et al.(2000)]{Gebhardt00} Gebhardt, K., Bender, R., Bower, G., et al.\ 2000,
 \apjl, 539, L13

\bibitem[Godet et al.(2012)]{godet12} Godet, O., Plazolles, B., Kawaguchi, T., et al.\ 2012, \apj, 752, 34 

\bibitem[Graham(2007)]{graham07} Graham, A.~W.\ 2007, \mnras, 379, 711 
\bibitem[Greene \& Ho(2004)]{greene04} Greene, J.~E., \& Ho, L.~C.\ 2004, \apj, 610, 722 
\bibitem[Greene et al.(2006)]{greene06} Greene, J.~E., Ho, L.~C., \& Ulvestad, J.~S.\ 2006, \apj, 636, 56  
%\bibitem[Greene \& Ho(2007b)]{greene07b} Greene, J.~E., \& Ho, L.~C.\ 2007b, \apj, 670, 92
\bibitem[Greene \& Ho(2007)]{greene07a} Greene, J.~E., \& Ho, L.~C.\ 2007a, \apj, 656, 84 
\bibitem[Greene \& Ho(2007b)]{greene07b} Greene, J.~E., \& Ho, L.~C.\ 2007b, \apj, 670, 92
\bibitem[Greene et al.(2008)]{greene08} Greene, J.~E., Ho, L.~C., \& Barth, A.~J.\ 2008, \apj, 688, 159-179  
\bibitem[Greene(2012)]{greene12} Greene, J.~E.\ 2012, Nature Communications, 3, 1304
\bibitem[G{\"u}ltekin et al.(2014)]{gultekin14} G{\"u}ltekin, K., Cackett, E.~M., King, A.~L., Miller, J.~M., \& Pinkney, J.\ 2014, \apjl, 788, L22 

\bibitem[Gunn et al.(1998)]{gunn98} Gunn, J.~E., Carr, M., Rockosi, C., et al.\ 1998, \aj, 116, 3040 

\bibitem[Heckman et al.(2005)]{heckman05} Heckman, T.~M., Ptak, A., Hornschemeier, A., \& Kauffmann, G.\ 2005, \apj, 634, 161 
\bibitem[Kormendy \& Ho(2013)]{kormendy13} Kormendy, J., \& Ho, L.~C.\ 2013, \araa, 51, 511 
\bibitem[Ho(1999)]{ho99} Ho, L.~C.\ 1999, \apj, 516, 672 
\bibitem[Ho et al.(2012)]{ho12} Ho, L.~C., Kim, M., \& Terashima, Y.\ 2012, \apjl, 759, L16 
\bibitem[Ho \& Kim(2016)]{ho16} Ho, L.~C., \& Kim, M.\ 2016, \apj, 821, 48 

\bibitem[Ivison et al.(2010)]{ivison10} Ivison, R.~J., Magnelli, B., Ibar, E., et al.\ 2010, \aap, 518, L31 
\bibitem[Jiang et al.(2011a)]{jiang11a} Jiang, Y.-F., Greene, J.~E., Ho, L.~C., Xiao, T., \& Barth, A.~J.\ 2011a, \apj, 742, 68
\bibitem[Jiang et al.(2011b)]{jiang11b} Jiang, Y.-F., Greene, J.~E., \& Ho, L.~C.\ 2011b, \apjl, 737, L45 
\bibitem[Jiang et al.(2013)]{jiang13} Jiang, N., Ho, L.~C., Dong, X.-B., Yang, H., \& Wang, J.\ 2013, \apj, 770, 3 
\bibitem[Jin et al.(2012a)]{jin12a} Jin, C., Ward, M., Done, C., \& Gelbord, J.\ 2012a, \mnras, 420, 1825 
\bibitem[Jin et al.(2012b)]{jin12b} Jin, C., Ward, M., \& Done, C.\ 2012b, \mnras, 425, 907 

\bibitem[Jin et al.(2016)]{jin16} Jin, C., Done, C., \& Ward, M.\ 2016, \mnras, 455, 691 

\bibitem[Kamizasa et al.(2012)]{kamizasa12} Kamizasa, N., Terashima, Y., \& Awaki, H.\ 2012, \apj, 751, 39
\bibitem[Kewley et al.(2006)]{kewley06} Kewley, L.~J., Groves, B., Kauffmann, G., \& Heckman, T.\ 2006, \mnras, 372, 961 
\bibitem[Kellermann et al.(1989)]{kellermann89} Kellermann, K.~I., Sramek, R., Schmidt, M., Shaffer, D.~B., \& Green, R.\ 1989, \aj, 98, 1195  
\bibitem[Kormendy \& Kennicutt(2004)]{kormendy04} Kormendy, J., \& Kennicutt, R.~C., Jr.\ 2004, \araa, 42, 603 
\bibitem[Laurikainen et al.(2014)]{laurikainen14} Laurikainen, E., Salo, H., Athanassoula, E., Bosma, A., \& Herrera-Endoqui, M.\ 2014, \mnras, 444, L80 

\bibitem[Leighly et al.(2007)]{leighly07} Leighly, K.~M., Halpern, J.~P., Jenkins, E.~B., et al.\ 2007, \apj, 663, 103 

\bibitem[Li \& Shen(2012)]{2012ApJ...757L...7L} Li, Z.-Y., \& Shen, J.\ 2012, \apjl, 757, L7 
\bibitem[Li et al.(2016)]{li16} Li, Y.-P., Yuan, F., \& Wang, Q.~D.\ 2016, arXiv:1611.02904 
\bibitem[Lin et al.(2013)]{lin13} Lin, D., Irwin, J.~A., Godet, O., Webb, N.~A., \& Barret, D.\ 2013, \apjl, 776, L10
\bibitem[Ludlam et al.(2015)]{ludlam15} Ludlam, R.~M., Cackett, E.~M., G{\"u}ltekin, K., et al.\ 2015, \mnras, 447, 2112 
\bibitem[McHardy et al.(2006)]{mchardy06} McHardy, I.~M., Koerding, E., Knigge, C., Uttley, P., \& Fender, R.~P.\ 2006, \nat, 444, 730 
\bibitem[Marconi \& Hunt(2003)]{marconi03} Marconi, A., \& Hunt, L.~K.\ 2003, \apjl, 589, L21 
\bibitem[Mahajan et al.(2010)]{mahajan10} Mahajan, S., Haines, C.~P., \& Raychaudhury, S.\ 2010, \mnras, 404, 1745 
\bibitem[Maraston(2005)]{2005MNRAS.362..799M} Maraston, C.\ 2005, \mnras, 362, 799 

\bibitem[McConnell \& Ma(2013)]{mcconnell13} McConnell, N.~J., \& Ma, C.-P.\ 2013, \apj, 764, 184 

\bibitem[Merloni et al.(2003)]{merloni03} Merloni, A., Heinz, S., \& di Matteo, T.\ 2003, \mnras, 345, 1057 

\bibitem[Moran et al.(1999)]{moran99} Moran, E.~C., Filippenko, A.~V., Ho, L.~C., et al.\ 1999, \pasp, 111, 801 

\bibitem[Moran et al.(2005)]{moran05} Moran, E.~C., Eracleous, M., Leighly, K.~M., et al.\ 2005, \aj, 129, 2108 

\bibitem[Moustakas et al.(2006)]{moustakas06} Moustakas, J., Kennicutt, R.~C., Jr., \& Tremonti, C.~A.\ 2006, \apj, 642, 775 
\bibitem[Miller et al.(2009)]{miller09} Miller, N.~A., Hornschemeier, A.~E., \& Mobasher, B.\ 2009, \aj, 137, 4436  
\bibitem[Miniutti et al.(2009)]{miniutti09} Miniutti, G., Ponti, G., Greene, J.~E., et al.\ 2009, \mnras, 394, 443
\bibitem[Miniutti et al.(2013)]{miniutti13} Miniutti, G., Saxton, R.~D., Rodr{\'{\i}}guez-Pascual, P.~M., et al.\ 2013, \mnras, 433, 1764  
%\bibitem[Mihos \& Hernquist(1994)]{1994ApJ...437L..47M} Mihos, J.~C., \& Hernquist, L.\ 1994, \apjl, 437, L47 
\bibitem[Nicastro(2000)]{nicastro00} Nicastro, F.\ 2000, \apjl, 530, L65 

\bibitem[Nyland et al.(2012)]{nyland12} Nyland, K., Marvil, J., Wrobel, J.~M., Young, L.~M., \& Zauderer, B.~A.\ 2012, \apj, 753, 103 

\bibitem[Moran et al.(2014)]{moran14} Moran, E.~C., Shahinyan, K., Sugarman, H.~R., V{\'e}lez, D.~O., \& Eracleous, M.\ 2014, \aj, 148, 136 
\bibitem[Page et al.(2012)]{page12} Page, M.~J., Brindle, C., Talavera, A., et al.\ 2012, \mnras, 426, 903 
\bibitem[Pan et al.(2015)]{pan15} Pan, H.-W., Yuan, W., Zhou, X.-L., Dong, X.-B., \& Liu, B.\ 2015, \apj, 808, 163 
\bibitem[Panessa et al.(2006)]{panessa06} Panessa, F., Bassani, L., Cappi, M., et al.\ 2006, \aap, 455, 173 
\bibitem[Peng et al.(2002)]{peng02} Peng, C.~Y., Ho, L.~C., Impey, C.~D., \& Rix, H.-W.\ 2002, \aj, 124, 266  
\bibitem[Peng et al.(2010)]{peng10} Peng, C.~Y., Ho, L.~C., Impey, C.~D., \& Rix, H.-W.\ 2010, \aj, 139, 2097 
\bibitem[Plotkin et al.(2016)]{plotkin16} Plotkin, R.~M., Gallo, E., Haardt, F., et al.\ 2016, \apj, 825, 139 

\bibitem[Pohlen \& Trujillo(2006)]{pohlen06} Pohlen, M., \& Trujillo, I.\ 2006, \aap, 454, 759 

\bibitem[Polletta et al.(2007)]{2007ApJ...663...81P} Polletta, M., Tajer, M., Maraschi, L., et al.\ 2007, \apj, 663, 81 
\bibitem[Proga(2005)]{proga05} Proga, D.\ 2005, \apjl, 630, L9 
\bibitem[Reines \& Deller(2012)]{reines12} Reines, A.~E., \& Deller, A.~T.\ 2012, \apjl, 750, L24 
\bibitem[Reines et al.(2013)]{reines13} Reines, A.~E., Greene, J.~E., \& Geha, M.\ 2013, \apj, 775, 11
\bibitem[Reis \& Miller(2013)]{reis13} Reis, R.~C., \& Miller, J.~M.\ 2013, \apjl, 769, L7 
\bibitem[Rossa et al.(2006)]{2006AJ....132.1074R} Rossa, J., van der Marel, R.~P., B{\"o}ker, T., et al.\ 2006, \aj, 132, 1074 6
\bibitem[Schramm et al.(2013)]{schramm13} Schramm, M., Silverman, J.~D., Greene, J.~E., et al.\ 2013, \apj, 773, 150 
\bibitem[Skrutskie et al.(2006)]{2006AJ....131.1163S} Skrutskie, M.~F., Cutri, R.~M., Stiening, R., et al.\ 2006, \aj, 131, 1163  
\bibitem[Schmidt et al.(1992)]{1992ApJ...398L..57S} Schmidt, G.~D., Stockman, H.~S., \& Smith, P.~S.\ 1992, \apjl, 398, L57
\bibitem[Seth et al.(2008)]{seth08} Seth, A., Ag{\"u}eros, M., Lee, D., \& Basu-Zych, A.\ 2008, \apj, 678, 116 
\bibitem[Seth et al.(2010)]{seth10} Seth, A.~C., Cappellari, M., Neumayer, N., et al.\ 2010, \apj, 714, 713 
\bibitem[Siegel et al.(2007)]{2007ApJ...667L..57S} Siegel, M.~H., Dotter, A., Majewski, S.~R., et al.\ 2007, \apjl, 667, L57 

\bibitem[Steffen et al.(2006)]{steffen06} Steffen, A.~T., Strateva, I., Brandt, W.~N., et al.\ 2006, \aj, 131, 2826 

\bibitem[Sun et al.(2013)]{sun13} Sun, L., Shu, X., \& Wang, T.\ 2013, \apj, 768, 167 (S13) 
\bibitem[Terashima et al.(2012)]{terashima12} Terashima, Y., Kamizasa, N., Awaki, H., Kubota, A., \& Ueda, Y.\ 2012, \apj, 752, 154 
\bibitem[Thornton et al.(2008)]{thornton08} Thornton, C.~E., Barth, A.~J., Ho, L.~C., Rutledge, R.~E., \& Greene, J.~E.\ 2008, \apj, 686, 892-910 

\bibitem[van Dokkum et al.(2015)]{vandokkum15} van Dokkum, P.~G., Abraham, R., Merritt, A., et al.\ 2015, \apjl, 798, L45 

\bibitem[Vasudevan \& Fabian(2009)]{vasudevan09} Vasudevan, R.~V., \& Fabian, A.~C.\ 2009, \mnras, 392, 1124 

\bibitem[Volonteri et al.(2003)]{volonteri03} Volonteri, M., Haardt, F., \& Madau, P.\ 2003, \apj, 582, 559 

\bibitem[Volonteri(2010)]{Volonteri10} Volonteri, M.\ 2010, \aapr, 18, 279 
\bibitem[Walcher et al.(2006)]{2006ApJ...649..692W} Walcher, C.~J., B{\"o}ker, T., Charlot, S., et al.\ 2006, \apj, 649, 692 
\bibitem[Wang et al.(2001)]{wang01} Wang, T.~G., Matsuoka, M., Kubo, H., Mihara, T., \& Negoro, H.\ 2001, \apj, 554, 233 
\bibitem[Wilkins et al.(2015)]{wilkins15} Wilkins, D.~R., Gallo, L.~C., Grupe, D., et al.\ 2015, \mnras, 454, 4440 

\bibitem[White \& Rees(1978)]{white78} White, S.~D.~M., \& Rees, M.~J.\ 1978, \mnras, 183, 341 

\bibitem[Wright et al.(2010)]{wright10} Wright, E.~L., Eisenhardt, P.~R.~M., Mainzer, A.~K., et al.\ 2010, \aj, 140, 1868-1881 

\bibitem[Wrobel \& Ho(2006)]{wrobel06} Wrobel, J.~M., \& Ho, L.~C.\ 2006, \apjl, 646, L95 

\bibitem[Xiao et al.(2011)]{xiao11} Xiao, T., Barth, A.~J., Greene, J.~E., et al.\ 2011, \apj, 739, 28 
\bibitem[Yuan et al.(2010)]{yuan10} Yuan, W., Liu, B.~F., Zhou, H., \& Wang, T.~G.\ 2010, \apj, 723, 508 
\bibitem[Yuan et al.(2014)]{yuan14} Yuan, W., Zhou, H., Dou, L., et al.\ 2014, \apj, 782, 55 
\bibitem[Zhou et al.(2015)]{zhou15} Zhou, X.-L., Yuan, W., Pan, H.-W., \& Liu, Z.\ 2015, \apjl, 798, L5 


\end{thebibliography}
\end{document}